\documentclass[aps,prb,twocolumn]{revtex4}
\usepackage{graphicx}
\newcommand{\ket}[1]{\vert #1 \rangle}
\newcommand{\abs}[1]{\vert #1 \vert}
\newcommand{\bra}[1]{\langle #1 \vert}
\begin{document}

\author{J.-B.\ Fouet}
\affiliation{Institut Romand de Recherche Num\'erique en Physique des
Mat\'eriaux (IRRMA), PPH-Ecublens, CH-1015 Lausanne, Switzerland}

\author{F.\ Mila}
\affiliation{Institute of Theoretical Physics, \'Ecole Polytechnique
F\'ed\'erale de Lausanne, CH-1015 Lausanne, Switzerland}

\author{D.\ Clarke}
\author{H.\ Youk}
\author{O.\ Tchernyshyov}
\affiliation{Department of Physics and Astronomy, Johns Hopkins University,
3400 North Charles Street, Baltimore, Maryland 21218, USA}

\author{P.\ Fendley}
\affiliation{Department of Physics, University of Virginia,
Charlottesville, Virginia 22904-4714, USA}
\author{R.\ M.\ Noack}
\affiliation{Fachbereich Physik, Philipps Universit\"at Marburg,
D-35032 Marburg, Germany}
\title{Condensation of magnons and spinons in a frustrated ladder}

\date{\today}

\begin{abstract}
Motivated by the ever-increasing experimental effort devoted to the
properties of frustrated quantum magnets in a magnetic field, we
present a careful and detailed theoretical analysis of a
one-dimensional version of this problem, a frustrated ladder with a
magnetization plateau at $m=1/2$. We show that even for purely
isotropic Heisenberg interactions, the magnetization curve exhibits
a rather complex behavior that can be fully accounted for in terms
of simple elementary excitations.  The introduction of anisotropic
interactions (e.g., Dzyaloshinskii-Moriya interactions) modifies
significantly the picture and reveals an essential difference
between integer and fractional plateaux. In particular, anisotropic
interactions generically open a gap in the region between the
plateaux, but we show that this gap closes upon entering fractional
plateaux. All of these conclusions, based on analytical arguments,
are supported by extensive Density Matrix Renormalization Group
calculations.

\end{abstract}

\maketitle

\section{Introduction}

Quantum phase transitions\cite{Sachdev-book} occur when the
variation of an external parameter (pressure, chemical composition
etc.) produces a singular change of a system's ground state.  If the
transition is continuous, the properties of the system in the
vicinity of the quantum critical point are dominated by universal
features, which can be described in field-theoretic
language.\cite{Sachdev-book} Quantum phase transitions and related
critical behavior are observed in a variety of physical systems,
from heavy-fermion compounds\cite{Stewart84} to quantum
magnets\cite{Bitko96} and cold atomic gases.\cite{Greiner02}

In antiferromagnets the external parameter is typically an applied
magnetic field ${\bf H} = (0,0,H)$.  If the longitudinal component
of the total spin, $S^z$, is a good quantum number, the magnetic
moment of the ground state $M = g \mu_B \langle S^z \rangle$ as a
function of $H$ may exhibit several plateaux, on which $dM/dH = 0$,
separated by regions of continuously varying magnetization.  The
ends of a plateau are quantum critical points separating an
incompressible ground state with an energy gap from a compressible
state with gapless excitations. Affleck\cite{Affleck91} noted in the
context of spin 1 chains that such a phase transition is similar to
the condensation in a system of interacting bosons, a point of view
reemphasized and extended in the context of coupled spin 1/2
ladders.\cite{giamarchi}  The magnetic field and magnetic moment
play the roles of the chemical potential and particle number.  The
condensing objects are magnons, the quasiparticles carrying spin
$\Delta S^z = \pm 1$.  The field theory describing the bosons near
the Bose-condensation point\cite{Fisher89} is equally applicable to
the end points of a magnetic plateau.

Not every magnetization plateau ends in a simple condensation of
magnons.  If the plateau state breaks a symmetry of the lattice while
the gapless state does not, the transition must restore the broken
lattice symmetry.  The universal properties of such a transition are
expected to be different.  In one spatial dimension, the picture based
on the magnon condensation is generally applicable to ``integer''
magnetization plateaux defined as follows:\cite{Oshikawa97a} the spin
per unit cell differs from the maximal value by an integer.  A
fractional magnetization plateau may end in a phase transition
belonging to a different universality class.  Examples of such
behavior were recently discussed by a number of
authors.\cite{Fendley04,Lecheminant04,Essler04}

In this paper, we present a theoretical study of quantum phase
transitions in a one-dimensional model antiferromagnet exhibiting both
integer and fractional magnetization plateaux.  We employ numerical
methods to observe critical behavior and compare the results to
predictions of the appropriate field theories.  The outcome of our
work stresses the importance of anisotropic interactions in the
vicinity of quantum critical points, a point raised previously by
previous authors.\cite{Xia88,Oshikawa97b} Even a weak anisotropy makes a
significant impact on the quantum phase transitions in question and,
furthermore, the effects vary substantially between different
universality classes.

The paper is organized as follows.  The model system and details of
the numerical method are described in Sec.~\ref{sec-description}.
In Sec.~\ref{sec-isotropic} we discuss the ground states and phase
transitions in the model with isotropic interactions.  The influence
of anisotropy is the subject of Sec.~\ref{sec-anisotropic}, and a summary
of the results is given in Sec.~\ref{sec-conclusion}.

\section{Description of the model}
\label{sec-description}

\subsection{Hamiltonian and ground states}

\begin{figure}
\includegraphics[width=0.95\columnwidth]{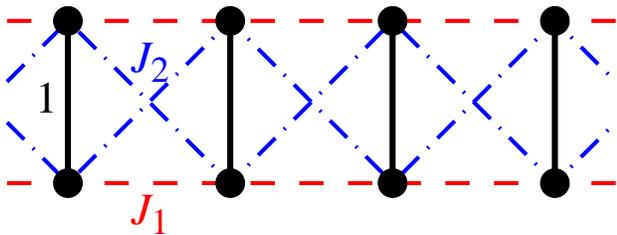}
\caption{(Color online) The frustrated ladder.
The exchange couplings are set to 1 on
the vertical rungs, $J_1 = 0.55$ on horizontal legs, and $J_2 = 0.7$
on the diagonals.  }
\label{fig-ladder}
\end{figure}

Our model system is the frustrated ladder with spin-$1/2$ spins and
with Heisenberg exchange.\cite{Mila98} The largest exchange
couplings, of strength 1, are on the rungs, with somewhat weaker
couplings $J_1$ along the legs and $J_2$ along plaquette diagonals,
see Fig.~\ref{fig-ladder}.  The Hamiltonian is
\begin{eqnarray}
\mathcal{H}&=&\sum_{n=1}^{L}
 [\mathbf{S}_{n,1}\cdot \mathbf{S}_{n,2}
 - H (S_{n,1}^z + S_{n,2}^z)]
\nonumber\\
&& + J_1 \sum_{n=1}^{L-1}
 (\mathbf{S}_{n,1}\cdot \mathbf{S}_{n+1,1}
  +\mathbf{S}_{n,2}\cdot \mathbf{S}_{n+1,2})
\label{eq-H1}\\
&& +J_2 \sum_{n=1}^{L-1}
 (\mathbf{S}_{n,1}\cdot \mathbf{S}_{n+1,2}
  +\mathbf{S}_{n,2}\cdot \mathbf{S}_{n+1,1}).
\nonumber
\end{eqnarray}
A ladder without diagonal links ($J_2=0$) was examined in
Ref.\ \onlinecite{Cabra97}.  The point $J_1=J_2$ is special: the spin of
each rung is a conserved quantity and the ground state is known
exactly.\cite{Honecker00} Conditions for the existence of a
$m=1/2$ plateau in the general case were given in
Ref.\ \onlinecite{Mila98}.

In this work, we study the case where the interrung couplings $J_1$ and
$J_2$ are nonvanishing and similar in strength.  Proximity to the
exactly solvable model $J_1=J_2$ suggests a partition of the Hamiltonian
into an exactly solvable part plus a small perturbation.  To that end,
it is convenient to introduce the operators of the total spin of a rung
$\mathbf{S}_n = \mathbf{S}_{n,1} + \mathbf{S}_{n,2}$ and the spin
difference $\mathbf{D}_n = \mathbf{S}_{n,1} - \mathbf{S}_{n,2}$.
The exchange part of the Hamiltonian is then a sum of three terms:
\begin{eqnarray}
\mathcal{H}_0 &=& \sum_{n=1}^{L} (\mathbf{S}_{n})^2/2 - H S_n^z,
\nonumber
\\
\mathcal{H}_1 &=& J
\sum_{n=1}^{L-1} \mathbf{S}_{n} \cdot \mathbf{S}_{n+1},
\label{eq-H}
\\
\mathcal{H}_2 &=& \frac{\delta J}{2}
\sum_{n=1}^{L-1} \mathbf{D}_{n} \cdot \mathbf{D}_{n+1},
\nonumber
\end{eqnarray}
where $J = (J_1 + J_2)/2$ and $\delta J = J_1 - J_2$.  The algebra of
$\mathbf{D}$ operators is discussed in Appendix \ref{ap-algebra}.

\begin{figure}
\includegraphics[width=0.95\columnwidth]{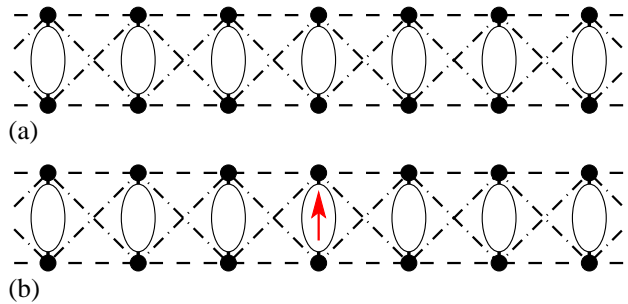}
\caption{(Color online)
(a) A sketch of the ground state at low magnetic fields: all
of the rungs are in the $S=0$ state.  (b) Elementary excitations out
of this state are magnons carrying spin $\Delta S^z = +1$. }
\label{fig-zero}
\end{figure}

The physics of the model is rather simple when the energy scales are
well separated,
\begin{equation}
\delta J \ll J \ll 1.
\label{eq-scales}
\end{equation}
The dominant term $\mathcal{H}_0$ favors a spin singlet on every rung
at low fields $H<1$ as depicted in Fig.~\ref{fig-zero}(a).  In high fields,
$H>1$, the state of lowest energy is the $S^z = +1$ component of the
triplet.  (The other two components of the triplet are high-energy
states in any magnetic field.)  The all-singlet and all-triplet states
are the origins of the two {\em integer} magnetization plateaux with
the spin per rung $m = S^z/L = 0$ and 1, respectively.

\begin{figure}
\includegraphics[width=0.95\columnwidth]{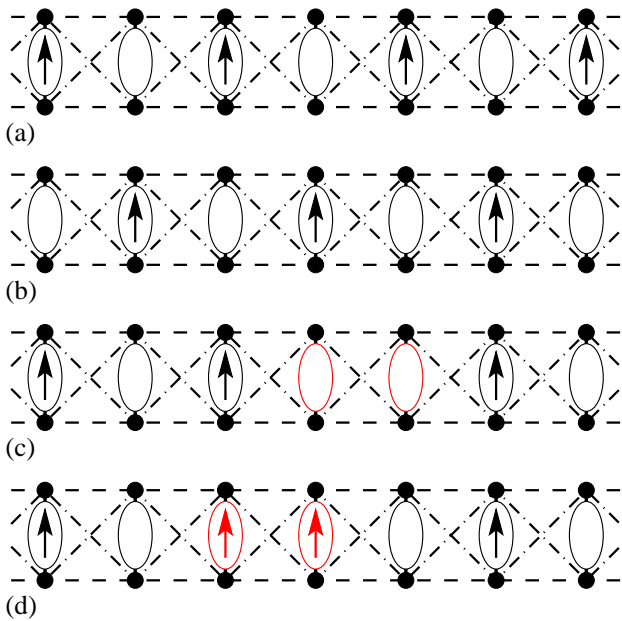}
\caption{(Color online)
(a) and (b) depict the two ground states of the fractional plateau with
$Z_2$ translational order.  Elementary excitations are domain walls
carrying $S^z = -1/2$ at (c) the low-field end of the plateau  and $S^z
= +1/2$ at (d) the high end.}
\label{fig-half}
\end{figure}

The next term $\mathcal{H}_1$ represents a repulsion between
triplets on neighboring rungs.  This repulsion does not allow the
triplets to condense on all rungs at once and introduces an
intermediate, {\em fractional} magnetization plateau with a triplet
on every other rung and spin per rung $m = 1/2$.  Accordingly, there
are two ground states breaking the translational symmetry,
illustrated in Figs.~\ref{fig-half}(a) and (b), which we refer to as
the N\'eel states.  The fractional plateau exists in the range of
fields $1 < H < 1 + 2J$.

Finally, the smallest term $\mathcal{H}_2$ endows the triplets with
mobility: unlike the total spin of the rung $\mathbf{S}_n$, the spin
difference $\mathbf{D}_n$ does not commute with the dominant term
$\mathcal{H}_0$.  The triplets acquire a kinetic energy of the order
$\delta J$.  As a result, the fractional and integer plateaux become
separated by gapless phases with a constantly varying magnetization
and with spin correlations that decay as a power of the distance
between sites.

The resulting Hamiltonian (\ref{eq-H}) has an axial O(2) symmetry.
However, in real materials that symmetry is often violated by small
additional interactions induced by the relativistic spin-orbital coupling.
In many cases such interactions act like a staggered transverse magnetic
field.\cite{Xia88,Oshikawa97b}  We consider the influence of such
additional terms in Sec.~\ref{sec-anisotropic}.

\subsection{Elementary excitations}

At the end of the $m = 0$ plateau, the elementary excitations are
individual triplets carrying spin $\Delta S^z = +1$
[Fig.~\ref{fig-zero}(b)].  Similarly, low-energy excitations near the
end of the $m=+1$ plateau are isolated singlets in the background of
$S^z = +1$ triplets.  They carry spin $\Delta S^z = - 1$.  We refer to
both of these excitations as {\em magnons.}

The low-energy excitations near the ends of the fractional plateau are
domain walls with spin $\Delta S^z = -1/2$ at the lower-field end and
$\Delta S^z = +1/2$ at the high-field end [Figs.~\ref{fig-half}(c)
and (d)].  They will be referred to as {\em spinons.}

The quantum phase transitions between the plateaux and gapless phases
are triggered by the condensation of these elementary excitations.

\subsection{Mapping onto an XXZ spin chain, hard-core bosons, and fermions}
\label{subsec-XXZ}

In the regime described by Eq.~(\ref{eq-scales}), each rung can be found
either in the
singlet or in the $S^z = +1$ triplet state. Effectively, we can
treat this as a system with only two states per site, using
perturbation theory in $\delta J$ to define a Hamiltonian that acts
on this reduced Hilbert space.

By identifying the singlet and the $S^z=1$ triplet states with the
spin-up and spin-down states, respectively, we map the ladder onto a
spin-$1/2$ XXZ antiferromagnetic chain with an easy-axis
anisotropy.\cite{Mila98} Defining the usual spin-1/2 matrices $s^x,$
$s^y$ and $s^z$ acting on the reduced Hilbert space, one has
\begin{eqnarray}
\mathcal{H}_{\mathrm{xxz}}
&=& \sum_{n=1}^{L-1}
[j_x (s_n^x s_{n+1}^x + s_n^y s_{n+1}^y)
+ j_z \, s_n^z s_{n+1}^z]
\nonumber\\
&&- \, H_\mathrm{edge} (s_1^z + s_L^z)/2
- H_\mathrm{xxz} \sum_{n=1}^L s_n^z \, ,
\label{eq-XXZ}
\end{eqnarray}
where
\begin{eqnarray}
&&j_x = \delta J + \mathcal{O}(\delta J^2), \quad
j_z = J + \mathcal{O}(\delta J^2),
\nonumber\\
&&H_\mathrm{xxz} = H - 1 - J  + \mathcal{O}(\delta J^2).
\label{eq-XXZ-couplings}
\end{eqnarray}
The additional magnetic field $H_\mathrm{edge} = J$ at the chain
ends breaks the symmetry between the two N\'eel states of this
effective spin chain.  The edge field plays a role in the formation
of the ground state at the fractional plateau, where singlets and
$S^z=+1$ triplets have comparable energies.

The XXZ chain is gapped in the antiferromagnetic regime
$j_z/\abs{j_x} > 1$ and $H_\mathrm{xxz}=0$, with zero magnetization
in the ground state.\cite{YangIII}  The presence of the energy gap
means that the magnetization remains exactly zero in a finite range
of fields $-H_\mathrm{min} <H_\mathrm{xxz}<H_\mathrm{min}$.  At $\pm
H_\mathrm{min}$ the energy gap vanishes and the spinons condense.
(The corresponding fields in the ladder corresponds to the edge of
the $M=1/2$ plateau and are called $H_{c3}$ and $H_{c2}$,
respectively).  As $|H_\mathrm{xxz}|$ is increased further, the
ground state becomes a sea of spinons with a growing magnetization.
The system in this regime is a Luttinger liquid with continously
varying critical exponents.\cite{Haldane81} Finally, at
$H_\mathrm{xxz}=\pm H_\mathrm{max}$ the magnetization of the XXZ
chain saturates (in the ladder, this corresponds to the saturation
field $H_{c4}$ and to the field $H_{c1}$ at the end of the m=0
plateau).  The critical fields of the XXZ chain are known
exactly:\cite{YangIII}
\begin{equation}
H_\mathrm{max} = j_z +\abs{j_x}, \quad H_\mathrm{min} =
\abs{j_x}\sinh{g}\sum_{n=-\infty}^\infty \frac{(-1)^n}{\cosh{ng}},
\label{eq-XXZ-Hc}
\end{equation}
where $\cosh{g} = j_z/|j_x|$.

The system can also be viewed as a one-dimensional hard-core Bose gas:
the singlet state of a rung is mapped onto an empty site, and a $S^z = +1$
triplet becomes an occupied site.  The bosons have hopping amplitude
$\delta J/2$, strong nearest-neighbor repulsion $J$, and chemical
potential $H_\mathrm{xxz}$.

In one dimension, the hard-core constraint can be removed by replacing
the bosons with spinless fermions.  This representation is
particularly convenient when the Bose system is nearly empty ($m \ll
1$) or nearly filled ($1-m \ll 1$).  In these limits, the short-range
repulsion between the fermions is largely suppressed by the Pauli
principle.
Near the fractional plateau, when $|m-1/2| \ll 1$, it is
convenient to represent the domain walls as spinless
fermions.\cite{Fowler78}

\subsection{Numerical work}

When the energy scales do not form the hierarchy of Eq.\
(\ref{eq-scales}), one must resort to numerical methods.
Nonetheless, the general picture painted above remains largely
intact and, furthermore, the critical behavior near the quantum
phase transitions is expected to be universal.  To verify this, we
have numerically determined the ground state, its magnetization, and
the energy gap in a ladder with coupling constants $J_1 = 0.55$
along the legs and $J_2 = 0.7$ along the plaquette diagonals.  The
ground and first excited states were determined by the DMRG
method.\cite{White92} We have worked with ladders with up to 200
rungs, and have used the so-called ``finite algorithm'' version of
the DMRG method.\cite{Noack05} The use of open boundary conditions
allows us to study oscillations of magnetization $\langle S^z_n
\rangle$ induced by the presence of the ends.

For the model with isotropic interactions, we have carried out two
sweeps and have retained $m=600$ states in the system block. The
typical weight of discarded density-matrix eigenvalues is of order
$10^{-12}$. We have performed a few calculations with 6 sweeps and
$m=1000$ states; the energy difference was of the order of
$10^{-7}$. We have calculated the ground state in each $S_z$ sector
in zero magnetic field for sizes up to $N=400$ and by comparing
energies in a field have deduced the global ground state as a
function of magnetic field.

For the model with a staggered field, $S_z$ is no longer a good
quantum number, and we need to make a calculation for each value of
the magnetic  field $H$. Fortunately, due to the opening of the gap,
far fewer states are needed: carrying out two sweeps and keeping
only $m=200$ states for up to $N=200$ sites and $m=400$ for larger
systems, the typical discarded weight was of order $10^{-11}$.

The other subtle issue when performing DMRG simulations is the
choice of boundary conditions. Since one cannot access all excited
states with the DMRG, choosing the appropriate boundary conditions
can be crucial to obtaining relevant information about the
excitation gap.

In the case of an isotropic chain, for which the gap is not an
issue, we have worked with open boundary conditions.  As a result,
in the $m=1/2$ plateau the ladder with an even number of rungs $L$
has a single spinon in the ground state (see section III.B.2 for
details).  The small magnetization step in the middle of the
fractional plateau occurs when the spinon changes its spin from
$\Delta S^z = -1/2$ to $+1/2$.

\begin{figure}
\includegraphics[width=0.95\columnwidth]{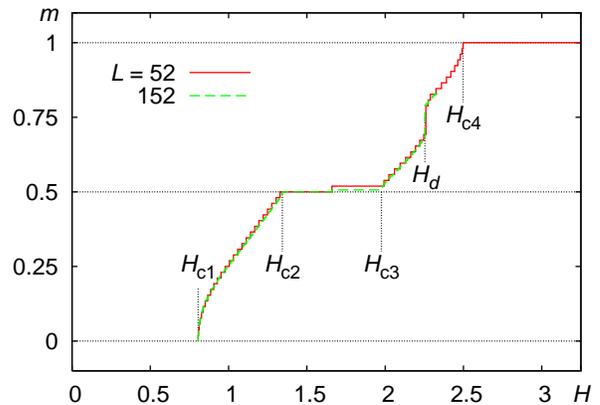}
\caption{(Color online)
Magnetization per rung $m = \langle S^z \rangle/L$.}
\label{fig-m}
\end{figure}

For systems with a staggered field, however, we have used asymmetric
boundary conditions, imposing different values of the rung couplings
at the edge: $J = 1.8$ for the first rung and  $J = 0.2$ for the
last one. These boundary conditions are necessary for the
determination of the energy gap in the formerly gapless regime
between $H_{c1}$ and $H_{c2}$ [Fig.~\ref{fig-m}].  With open
boundary conditions, the system has two nearly degenerate ground
states.  Altering the end rungs as described above lifts this
near-degeneracy and pushes one of the ground states well above the
first excited state.

\section{Ladder with isotropic interactions}
\label{sec-isotropic}

In this section, we present a detailed analysis of the properties of
the frustrated ladder without anisotropic interactions. The basic
properties of this system have been described
before.\cite{Honecker00,Cabra97,Mila98} In particular, the
magnetization curve is expected to have plateaux at $m=0$, $m=1/2$,
and $m=1$, to increase monotonically between the plateaux, and to
have square-root singularities at all plateau ends.  In practice,
the magnetization curve differs from the naive expectations in a
number of ways.  For example, the square-root singularity in the
magnetization near the fractional plateau is barely detectable; a
jump in magnetization is seen in the gapless region between the
$m=1/2$ and $m=1$ plateaux; the magnon mass near $H_{c1}$ differs
significantly from the prediction of perturbation theory.  We
discuss the main features of the magnetization curve below.

\subsection{Critical points}

The results for $L = 52$ and 152 rungs are shown in Fig.~\ref{fig-m}.
As expected, the magnetization curve has three plateaux at $m = 0$,
1/2, and 1.  The $m=0$ plateau ends at $H_{c1} = 0.806$, the $m=1/2$
plateau lies between $H_{c2} = 1.345$ and $H_{c3} = 1.974$, and a
fully polarized state with $m=1$ begins at $H_{c4} = 2.500$.  A finite
jump in the magnetization $\Delta m = 0.099$ is observed in a
compressible regime at $H_d = 2.254$.  We next discuss the critical fields.

\subsubsection{Magnon condensation at $H_{c4}$}

It is straightforward to calculate the end point of the fully
polarized state $m=1$.  When the energy scales separate well, i.e.,
in the regime described by Eq. (\ref{eq-scales}), the excitations with
the lowest energies are singlets with the dispersion
\begin{equation}
\epsilon_{0,0}(k) = H - 1 - 2J + \delta J \cos{k}.
\label{eq-e00}
\end{equation}
The singlets are gapped in fields $H > H_{0,0} = 1 + 2J + |\delta J|
= 2.4$ for our choice of couplings.  However, the numerics show that
the condensation occurs at a somewhat higher field.  This
discrepancy can be traced to a poor separation of energy scales:
$J = 0.625$ is not that small.  As a result, the first excitations
to condense are the $S^z = 0$ components of the triplet, whose
energy dispersion is
\begin{equation}
\epsilon_{1,0}(k) = H - 2J + 2J \cos{k} \, ,
\label{eq-e10}
\end{equation}
and the condensation field $H_{1,0} = 4J = 2.5$, in perfect agreement
with the numerics.

\subsubsection{Spinon condensation at $H_{c2}$ and $H_{c3}$}

To lowest order in $\delta J$, the end points of the fractional
plateau $m=1/2$ can be obtained in a similar way.  The spinons
carrying $S^z = \pm 1/2$ have energies
\begin{eqnarray}
\epsilon_{-1/2}(k) &=& (H-1)/2 + \delta J \cos{2k} + \mathcal{O}(\delta J^2),
\label{eq-spinon-1}\\
\epsilon_{+1/2}(k) &=& (1-H)/2 + J + \delta J \cos{2k}
+ \mathcal{O}(\delta J^2).
\nonumber
\end{eqnarray}
These spinons condense at
\begin{eqnarray}
H_{c2} &\approx& 1 + 2|\delta J| = 1.3 \ \ \mbox{and}
\label{eq-Hc23-1st}\\
H_{c3} &\approx& 1 + 2J -2|\delta J| = 1.95, \nonumber
\end{eqnarray}
respectively,
which are not that far off from the  values $H_{c2} = 1.345$
and $H_{c3} = 1.974$ obtained from the DMRG.

To improve the lowest-order estimate, we have expanded the parameters of
the XXZ chain (\ref{eq-XXZ-couplings}) and its critical field
$H_\mathrm{min}$ (\ref{eq-XXZ-Hc}) to $\mathcal{O}(\delta J^2)$.  The
physical origin of these corrections can be traced to quantum
fluctuations of spins in the ground and excited states of the $m=1/2$
plateau and its excitations.  The term
${\cal H}_2$
in the Hamiltonian connects the $S^{z}=-1/2$
spinon [two adjacent singlets, see Fig.~\ref{fig-half}(c)] to high-energy
states, shifting its energy by an amount $C \, \delta J^2 < 0$.  The
energy of the $S^{z}=+1/2$ spinon [Fig.~\ref{fig-half}(d)] is
unaffected by quantum fluctuations at this order.

The energy shift of the $S^z=-1/2$ spinon can be taken into account by
adding the following term to the Hamiltonian of the XXZ chain:
\begin{eqnarray}
\Delta \mathcal{H}_\mathrm{XXZ} &=&
\sum_{n} (1/2-s^z_n)(1/2-s^z_{n+1}) C \, \delta J^2
\\
&=& \mbox{const} - C \, \delta J^2 \sum s^z_n
+ C \, \delta J^2 \sum s^z_n s^z_{n+1}.
\nonumber
\end{eqnarray}
It can be seen that the added term affects both the Ising coupling
and the effective field of the XXZ chain:
\begin{eqnarray}
&&j_x = \delta J + \mathcal{O}(\delta J^3), \quad
j_z = J + C \, \delta J^2 + \mathcal{O}(\delta J^3),
\nonumber\\
&&H_\mathrm{xxz} = H - 1 - J + C \, \delta J^2 + \mathcal{O}(\delta J^3).
\label{eq-XXZ-couplings-2}
\end{eqnarray}

Expansion of Eq. (\ref{eq-XXZ-Hc}) in powers of $\delta J$ yields
the lower critical field of the XXZ chain
\begin{equation}
H_\mathrm{min} = J - 2 |\delta J| + (C + 1/2J) \delta J^2
+ \mathcal{O}(\delta J^3).
\end{equation}
The resulting condensation point of the $S^z=+1/2$ spinons
\begin{equation}
H_{c3} \approx 1 + 2J - 2 |\delta J| + \delta J^2/2J = 1.968
\end{equation}
is now very close to the DMRG value of 1.974.

The critical field of the $S^z=-1/2$ spinons is sensitive to the
energy shift $C \delta J^2$.  The function $C(J)$ is computed in
Appendix \ref{app-XXZ2}.  For $J=0.625$ we obtain $C = -2.026$ and thus
\begin{equation}
H_{c2} \approx 1 + 2 |\delta J| - (2C + 1/2J)\delta J^2 = 1.374.
\end{equation}
Comparing it to the DMRG value of 1.345, we see only a modest
improvement over the first-order result (\ref{eq-Hc23-1st}).  The
apparent reason for the slow convergence of the perturbation series
for $H_{c2}$ is the fairly small energy gap ($\Delta = 0.255$)
separating the $S^z=-1/2$ spinons from higher-energy states (see
Appendix \ref{app-XXZ2}).

\subsubsection{Magnetization jump at $H_d = 2.254$}

Between $H_{c3}$ and $H_{c4}$, the ground state switches from a
mixture of $S^z=+1$ triplets and singlets to one of $S^z=+1$ and
$S^z=0$ triplets.  This transition is accidental in the sense that it
is not accompanied by any change in symmetry.  It is therefore not
surprising that the change is accompanied by a discontinuity in the
magnetization.

\begin{figure}
\includegraphics[width=0.95\columnwidth]{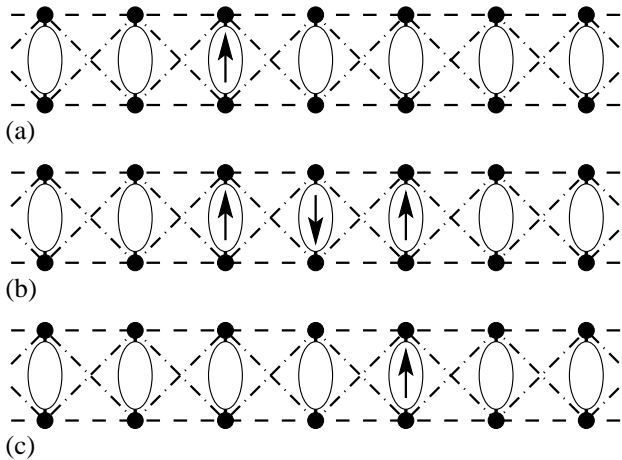}
\caption{Corrections to the kinetic energy of a magnon at the order
$\delta J^2$: (a)$\rightarrow$(b) A pair of triplet states with net
spin zero is created adjacent to an existing magnon.
(b)$\rightarrow$(c) One of those triplet states annihilates with the
original magnon, leaving a new magnon on a different lattice site.
This process provides an additional channel for magnon motion and
thus lowers the effective magnon mass.} \label{fig-2nd}
\end{figure}

\subsubsection{Magnon condensation at $H_{c1}$}

The energy of an isolated $\Delta S^z = 1$ excitation at the $m=0$ plateau is
\begin{equation}
\epsilon_{1,1}(k) = 1 - H + \delta J \, \cos{k}
+ \mathcal{O}(\delta J^2).
\end{equation}
Thus, to first order in $\delta J$, the magnon condensation is
expected at $H_{c1} = 1 - |\delta J| = 0.85$, not very far from the
DMRG result 0.806.

At $\mathcal{O}(\delta J^2)$ quantum fluctuations not only lower the
energy of the magnon but also change its hopping amplitude
(Fig.~\ref{fig-2nd}).  The magnitude of the second-order correction is
rather large, again thanks to a small energy gap ($\Delta = 0.125$)
between the magnon and higher-energy states.  See Appendix
\ref{ap-mass} for details.

\subsection{Magnetization patterns}

Thanks to the presence of an energy gap, both the total spin $S^z$ and
the average spin of an individual rung $\langle S^z_n \rangle$ remain
exactly zero in the low-field regime $|H| < H_{c1}$.  As the field
increases beyond $H_{c1}$, both the local and the total spin begin to
increase.  In a finite ladder, the local magnetization $\langle S^z_n
\rangle$ is distributed in a nonuniform way, revealing interference
patterns.

\subsubsection{$m \ll 1$: dilute magnons}
\label{section-isotropic-Hc1}

\begin{figure}
\includegraphics[width=0.95\columnwidth]{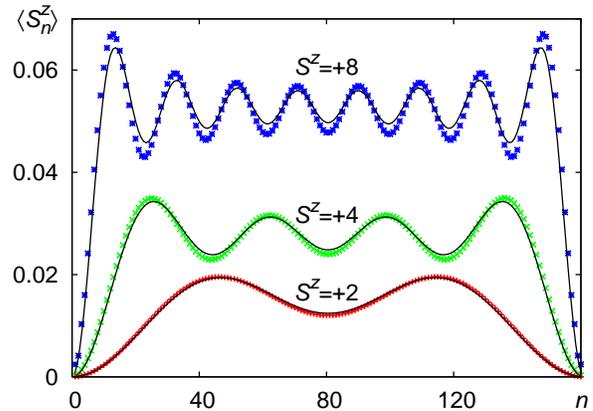}
\caption{(Color online)
Distribution of spin $\langle S^z_n \rangle$ as a function of
rung position $n$ in the ground state off of
the $m=0$ plateau.  The
symbols are the numerical results at $S^z = 2$, 4, and 8.  The curves
are spin distributions for 2, 4, and 8 magnons treated as free
fermions carrying spin $\Delta S^z = +1$.  Calculations are on a
ladder with $L = 160$ rungs. }
\label{fig-magnons}
\end{figure}

At low concentrations, the magnetization is carried by individual
magnons which can be viewed as hard-core bosons or, more
conveniently, as free fermions with spin $\Delta S^z = +1$.  (The
nearest-neighbor repulsive potential acting between magnons is
rendered irrelevant by the Fermi statistics.)  Treating the magnons
as ideal fermions one obtains a magnetization distribution
\begin{equation}
\langle S^z_n \rangle = \sum_{k=1}^{S^z} |\psi_k(n)|^2,
\label{eq-Sz-magnons}
\end{equation}
where, in the continuum approximation, $\psi_k(n) = \sqrt{2/L} \,
\sin{(\pi k n/L)}$ is the wavefunction of a nonrelativistic fermion in
a box of length $L+1$.  This simple model agrees well with the
magnetization distribution obtained numerically at small values of the
magnetization $m = S^z/L$ (Fig.~\ref{fig-magnons}).  Deviations
already become
noticeable when $m$ reaches 1/20.

\begin{figure}
\includegraphics[width=0.95\columnwidth]{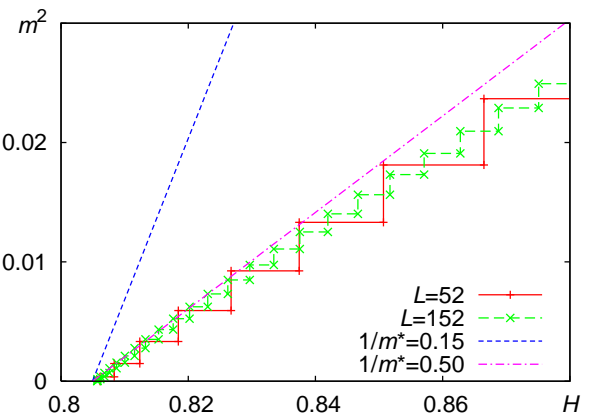}
\caption{(Color online)
The square of magnetization density $m$ off the $m=0$ plateau.}
\label{fig-m2detail}
\end{figure}

The free-fermion approach predicts a fast initial growth of
magnetization $m = S^z/L$:
\begin{equation}
m \sim \Delta S^z \, k_F/\pi
= \pi^{-1} \, |\Delta S^z|^{3/2} \, \sqrt{2 m^* |H-H_{c1}|},
\label{eq-m-Hc1}
\end{equation}
where the inverse mass $1/m^* \approx |\delta J|$.  We have found
numerically that $m^2$ indeed rises linearly with $H$
(Fig.~\ref{fig-m2detail}).  However, the slope $d m^2/dH$ is less
than a third of the calculated value of $2\pi^{-2}|\delta J|^{-1}$.

To track down the source of the discrepancy, we have computed the
contribution of higher-order processes to the magnon dispersion. The
$\mathcal O(\delta J^2)$ term turns out to be {\em larger} than the
first-order result.  This can be understood on a qualitative level
by considering a typical $\mathcal O(\delta J^2)$ contribution to
the kinetic energy of the magnon in which two additional magnons are
created and destroyed (Fig.~\ref{fig-2nd}).  The lowest-lying
3-magnon state has an energy of $\Delta = 2-3J$ above that of a
single magnon.  For $J = 0.625$, this energy gap $\Delta = 0.125$
is comparable to the small parameter $\delta J = -0.15$, meaning that
the lowest-order perturbation theory in $\delta J$ may not give
reliable results.  See Appendix \ref{ap-mass} for details.

\subsubsection{$|m-1/2| \ll 1$: dilute spinons}
\label{section-isotropic-Hc2}

\begin{figure}
\includegraphics[width=0.95\columnwidth]{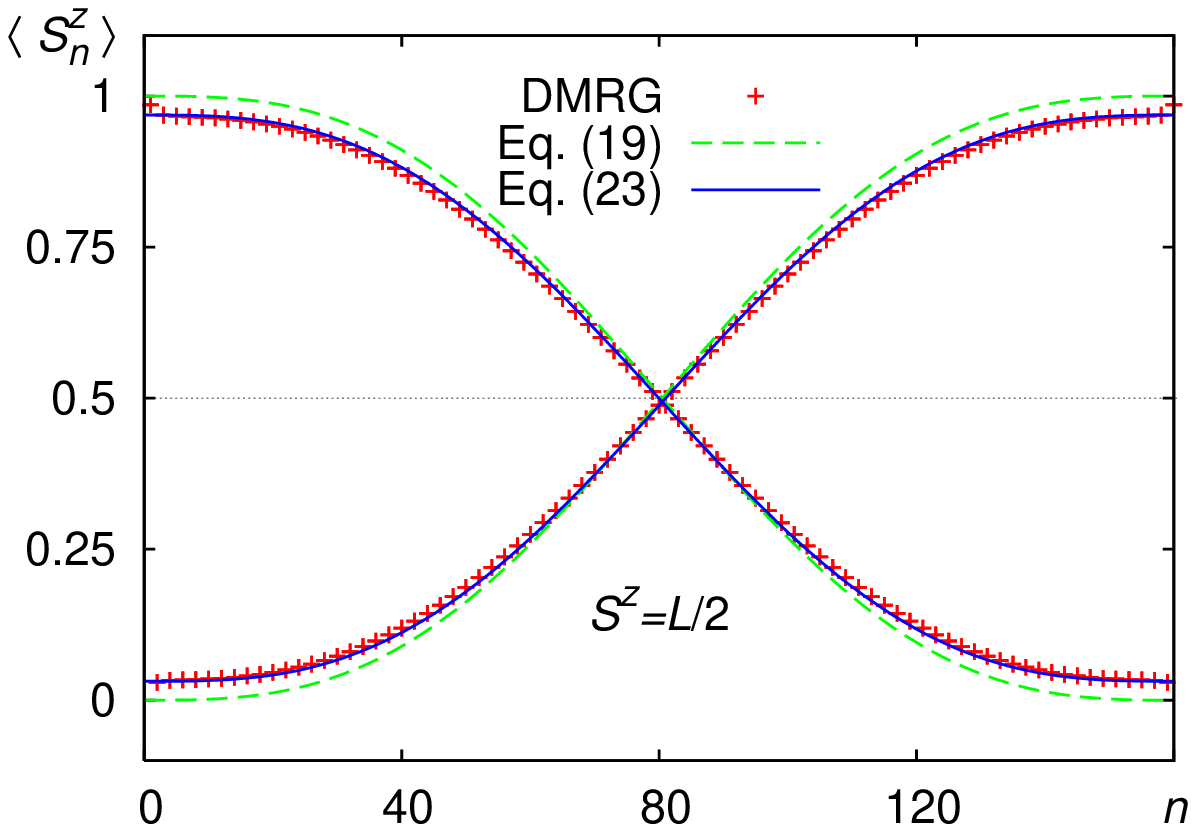}
\includegraphics[width=0.95\columnwidth]{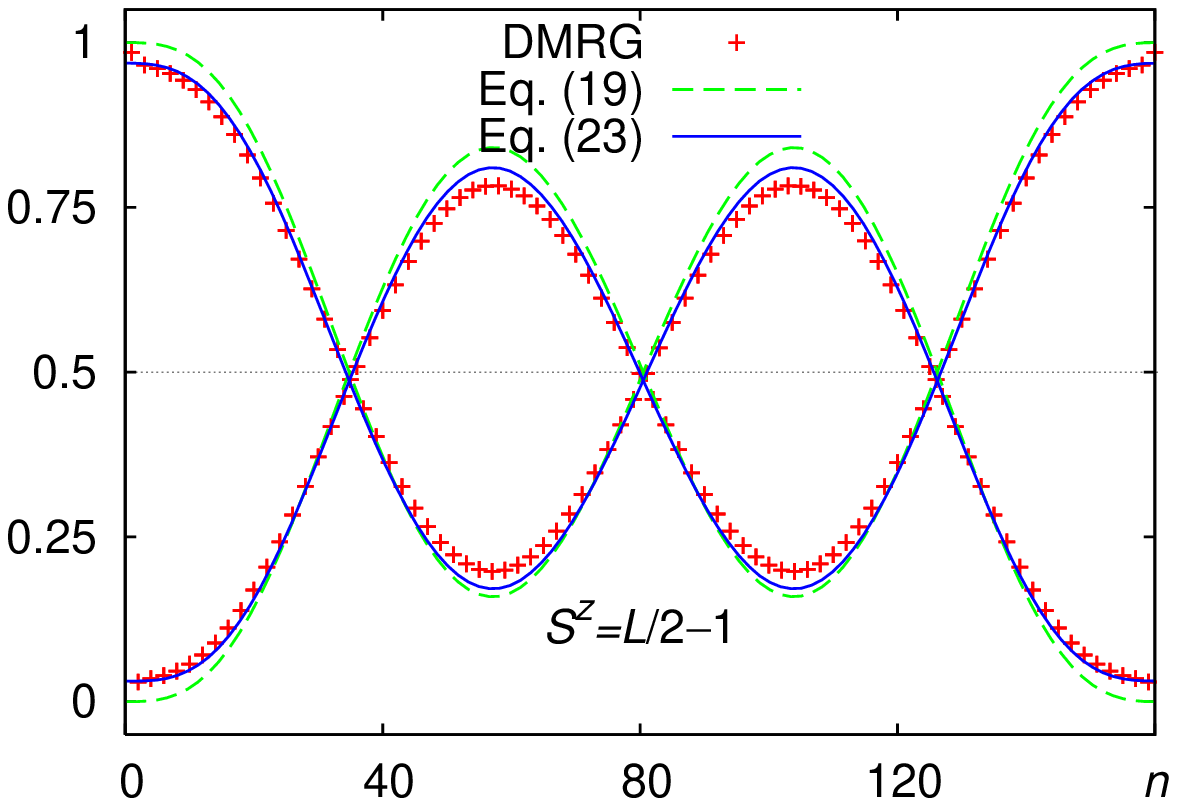}
\caption{(Color online)
Distribution of magnetization $\langle S^z_n \rangle$ in a
ladder with $L=160$ rungs (a) Exactly at the magnetization plateau
$m=1/2$, or $S^z = L/2$, the ladder contains one spinon.  (b) At $S^z
= L/2-1$, three spinons are present.}
\label{fig-spinon1}
\end{figure}

The numerically determined distribution of magnetization $\langle
S^z_n \rangle$ exactly at $m = 1/2$ ($S^z = 80$ in a ladder with
$L=160$ rungs) is shown in Fig.~\ref{fig-spinon1}(a).  Contrary to our
initial expectations, the plateau state is {\em not} a simple N\'eel
state with a constant staggered magnetization on top of a constant
background $1/2 + (-1)^n/2$.  The staggered magnetization is highly
nonuniform.

\begin{figure}
\includegraphics[width=0.95\columnwidth]{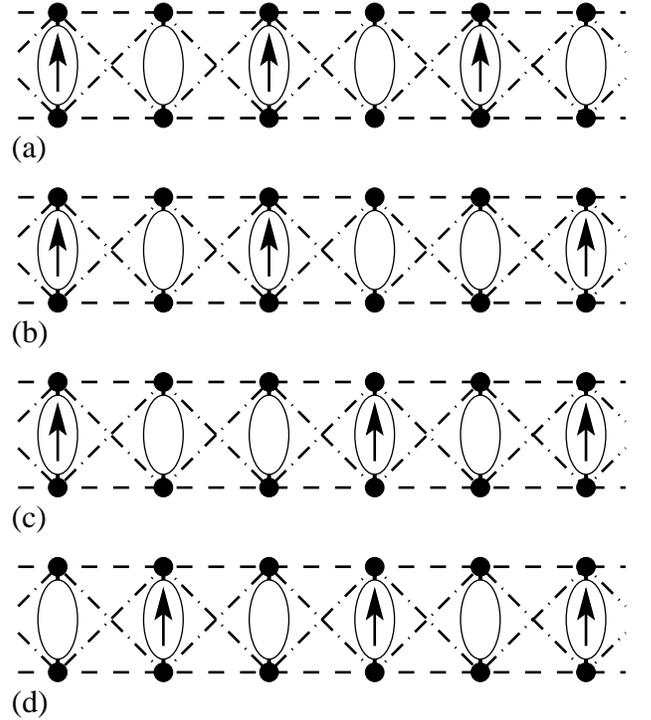}
\caption{Low-energy states of a ladder with an even number of rungs
at magnetization density $m=1/2$.}
\label{fig-1-spinon}
\end{figure}

To understand this result, consider the $m=1/2$ plateau state in a
finite ladder with an even number of rungs $L$.  To
$\mathcal{O}(\Delta J^0)$, its ground states are configurations with
$L/2$ triplets with no two triplets next to each other.  There are
only two such states in a ladder with periodic boundary conditions.
In contrast, for open boundary conditions there are $L/2+1$ such
configurations: two N\'eel states and $L/2-1$ states with a
singlet-singlet domain wall (Fig.~\ref{fig-1-spinon}).  The term
$\mathcal{H}_2$ delocalizes the domain wall and is ultimately
responsible for the strong modulation of the staggered
magnetization.

The mapping onto the XXZ chain provides an alternative perspective.
Ordinarily, the two N\'eel states of the XXZ chain are degenerate
even in the presence of a uniform magnetic field.  However, our
system (\ref{eq-XXZ}) has an extra magnetic field of strength $-J/2$
at the ends, so that both end spins prefer the $+1/2$ state.  In a
chain of even length, this inevitably leads to the formation of a
domain wall.  The cost of the wall ($+J/2$) is exactly offset by the
reduction of the energy of the end spin ($-J/2$); a negative
delocalization energy of the domain wall ($-|\delta J|$) lowers the
energy of this state relative to the N\'eel state.  Thus the
effective XXZ chain always has spins $+1/2$ at the ends in this
regime.  Accordingly, the end rungs of the ladder are in the triplet
state and thus there must be a single domain wall somewhere in the
chain.

The strong modulation of the staggered magnetization evident in
Fig.~\ref{fig-spinon1}(a) can be traced to the delocalization of the
spinon.  Rung $n$ is in the state with spin $S = 1/2 - (-1)^n/2$ if
the domain wall is on its right, otherwise it has spin $S = 1/2 +
(-1)^n/2$.  For a rung near the left edge, the domain wall is almost
always on the right and vice versa.  Thus we expect
\begin{equation}
\langle S^z_n \rangle = \frac{1}{2}  - \frac{(-1)^n E(n)}{2}.
\label{eq-E0}
\end{equation}
where the smooth envelope $E(n)$ interpolates between $+1$ and $-1$
and has a node in the middle.

To evaluate the envelope $E(n)$, we adopt the continuum
approximation and treat the spinon as a nonrelativistic particle in
a one-dimensional box. Doing so yields
\begin{equation}
E(n) = \int_0^L \mathrm{sgn}(x-n) \, |\psi_1(x)|^2 \, dx
= 1 - \frac{2n}{L} + \frac{\sin{(2\pi n/L)}}{\pi}\, ,
\end{equation}
where $\psi_1(n) = \sqrt{2/L} \sin{(\pi n/L)}$ is the ground-state
wavefunction of a spinon.
The result is shown in Fig.~\ref{fig-spinon1}(a) as a dashed curve.
While the agreement is already rather good, further improvements can
be made, as explained below.

Just to the left of the $m=1/2$ plateau, the ladder contains a few
spinons with spin $\Delta S^z = -1/2$ each.  Because the first and
last rungs still remain in the triplet state, the ground state of a
ladder with an even number of rungs $L$ and spin $S^z = L/2 - p$
contains $r = 2p+1$ spinons.  At low concentrations $r/L$ the
spinons can be treated as ideal fermions.\cite{Fowler78} The
envelope of the staggered magnetization $E(n)$ is the expectation
value $\langle (-1)^{s(n)} \rangle$, where $s(n)$ is the number of
spinons to the left of rung $n$.  The average is taken over the
ground state of nonrelativistic fermions occupying the first $r$
levels with wavefunctions $\psi_k(n) = \sqrt{2/L}\sin{(\pi k n/L)}$.
The result for $r$ spinons is
\begin{equation}
E(n) = \det{S(n)}.
\label{eq-E-S}
\end{equation}
The $r \times r$ matrix $S(n)$ has elements
\begin{equation}
S_{ij}(n) = \int_0^L \mathrm{sgn}(x-n) \, \psi_i^*(x) \psi_j(x) \, dx,
\label{eq-S}
\end{equation}
where $\psi_i(x)$ are the spinon wavefunctions of the {\em occupied}
states: $i = 1 \ldots r$.  See Appendix \ref{ap-det} for a derivation.
The numerical data and the theoretical curve for $S^z = L/2 - 1$ (3
spinons) are shown in Fig.~\ref{fig-spinon1}(b).

The agreement between the theoretical curve and the numerical data can
be further improved by taking into account the spin $\Delta S^z =
-1/2$ carried by the domain walls,
\begin{equation}
\langle S^z_n \rangle = \frac{1}{2}  - \frac{(-1)^n E(n)}{2}
- \frac{1}{2}\sum_{k=1}^{r} |\psi_k(n)|^2,
\end{equation}
and by figuring in the reduction of staggered magnetization by quantum
fluctuations.  The leading effect is a virtual exchange of a singlet
and triplet on neighboring rungs.  This process increases the energy
by $J$ (the strength of triplet repulsion), has the matrix element
$\delta J/2 \ll J$, and thus can be treated as a small perturbation.
To a leading order in $\delta J/J$ we find a simple reduction of the
envelope $E(n)$ by the factor $1-(\delta J/J)^2$.  See Appendix
\ref{app-reduction} for details.

Close to the plateau when the spinon gas is dilute, the deviation of
the magnetization density from $1/2$ is expected to be proportional to
$|H-H_{c2}|^{1/2}$, in complete analogy to the magnon case
(\ref{eq-m-Hc1}).  However, the magnetization curve (Fig.~\ref{fig-m})
evidently remains linear almost all the way to $H_{c2}$, with only a
hint of an upturn right next to the plateau.  A possible reason for
this behavior could be the narrowness of the critical region near
$H_{c2}$ where the spinons can be treated as noninteracting fermions.
One argument in favor of this interpretation is the relative smallness
of the square-root term whose amplitude is $\pi^{-1} \, |\Delta
S^z|^{3/2} \, |2 m^*|^{1/2}$.  In comparison to the magnons, the
spinons have a reduced spin $\Delta S^z$ and a smaller mass $m^*$
(by a factor of 4 to leading order in $\delta J$).  We also note that
magnetization curves of an easy-axis XXZ chain\cite{YangIII}
show a similar trend: the square-root term near the N\'eel-ordered
state is relatively small.

\subsubsection{$0 < m < 1$: a Luttinger liquid}
\label{section-isotropic-gapless}

The gapless phase in the field range $H_{c1} < H < H_{c2}$ is a
Luttinger liquid\cite{Haldane81} with continuously varying critical
indices.  The critical properties of the ladder are expected to be
similar to those of the easy-axis XXZ chain in the gapless regime
$H_\mathrm{min} < |H_\mathrm{xxz}| < H_\mathrm{max}$ in which the
compressibility exponent $K$ decreases monotonically between 1 (dilute
magnons) and 1/4 (dilute spinons).\cite{Haldane80}

\begin{figure}
\includegraphics[width=0.95\columnwidth]{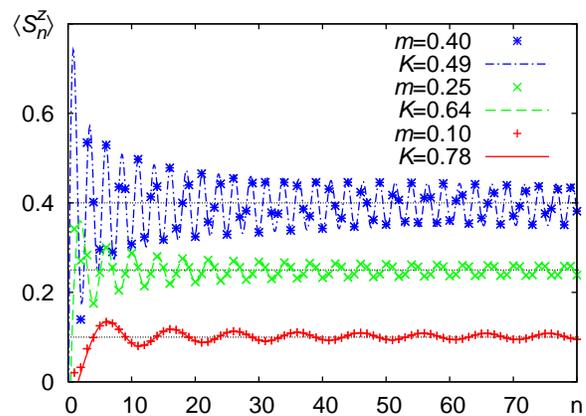}
\caption{(Color online)
Friedel oscillations of the local magnetization $\langle
S^z_n \rangle $ for three values of average magnetization $m$.  The
lines are best fits to the theoretical curves (\ref{eq-friedel}).  The
fits were performed in the range $n_0 \leq n \leq L+1-n_0$ with $n_0 =
5$.}
\label{fig-friedel}
\end{figure}

\begin{figure}
\includegraphics[width=0.95\columnwidth]{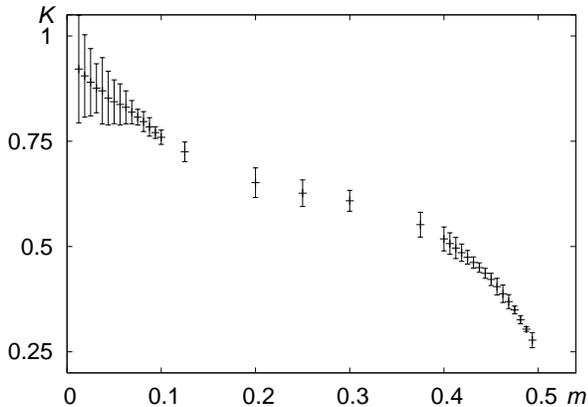}
\caption{The compressibility exponent $K$ as a function of magnetization $m$
in the gapless phase between fields $H_{c1}$ and $H_{c2}$.  The error
bars reflect the range of $K$ values obtained for different choices of the
short-range cutoff $n_0$.}
\label{fig-K}
\end{figure}

The compressibility exponent can be determined by examining the
Friedel oscillations in local magnetization $\langle S^z_n \rangle$
induced by the presence of the ends.\cite{White02} In a ladder of
length $L$, the leading behavior of magnetization away from the ends is
\begin{equation}
\langle S^z_n \rangle \sim
\mathrm{const} + \frac{a \cos{(2\pi m n + \beta)}}{[L\sin{(\pi n/L)}]^K},
\label{eq-friedel}
\end{equation}
where $m$ is the concentration of magnons and $\beta$ is a phase shift.
Fits of the numerical data to this form are shown in Fig.~\ref{fig-friedel}
for a ladder of length $L=160$.  The extracted exponent $K$ is shown in
Fig.~\ref{fig-K} as a function of the magnetization $m$.  While there is
a weak dependence on the short-range cutoff $n_0$, the trend is consistent
with a monotonic decrease of $K$ from 1 to 1/4.

\section{Influence of anisotropy}
\label{sec-anisotropic}

Experimentally accessible spin systems are almost inevitably
anisotropic.  While an anisotropy may be small numerically, it may
induce qualitative changes in the system's behavior because it lowers
the symmetry.  This is particularly important in the vicinity of a
critical point or phase: the presence of a symmetry-breaking term can
change the nature of a phase transition or even completely eliminate
it.  While the physical causes of anisotropies can vary, close to a
critical point their influence on the system can be expressed in terms
of a few relevant physical variables.  A well-known example is the
breaking of the axial O(2) symmetry of a spin chain under an external
magnetic field by the anisotropic Dzyaloshinskii-Moriya (DM)
interaction.  The action of the DM term is equivalent to that of a
weak staggered field transverse to the applied one.\cite{Xia88,Oshikawa97b}

A staggered field is uniquely defined in bipartite antiferromagnets.
A frustrated antiferromagnet can be partitioned into two sublattices
in more than one way.  Accordingly, several staggered fields can be
introduced in such cases.\cite{Sachdev92}  Three staggered fields are
potentially relevant to our ladder:
\begin{equation}
V = \sum_{n=1}^L \left[
h_{\pi0} (-1)^n S^x_n + h_{0\pi} D^x_n + h_{\pi\pi} (-1)^n D^x_n
\right]
\label{eq-staggered}
\end{equation}
(The fields are labeled by their Fourier indices.)  The most relevant
perturbation at a phase transition is that which couples directly to
an order parameter.  In our case ($J_1 < J_2 < J$), the ``condensate''
at $H_{c1}$ has the wavevector $(0,\pi)$.  The staggered field
$h_{0\pi}$ coupled to it breaks down the O(2) symmetry completely.  In
its presence, the quantum phase transition at $H_{c1}$ is expected to
become a crossover.  The other two staggered fields couple to the
square of the order parameter, generating an easy-axis anisotropy along
either $x$ or $y$ directions and thus lowering the symmetry from O(2)
down to $Z_2$.  In the absence of the more relevant staggered field
$h_{0\pi}$, the critical point at $H_{c1}$ will be preserved, but the
universality class is expected to change from
commensurate-incommensurate to Ising.

\begin{figure}
\includegraphics[width=0.95\columnwidth]{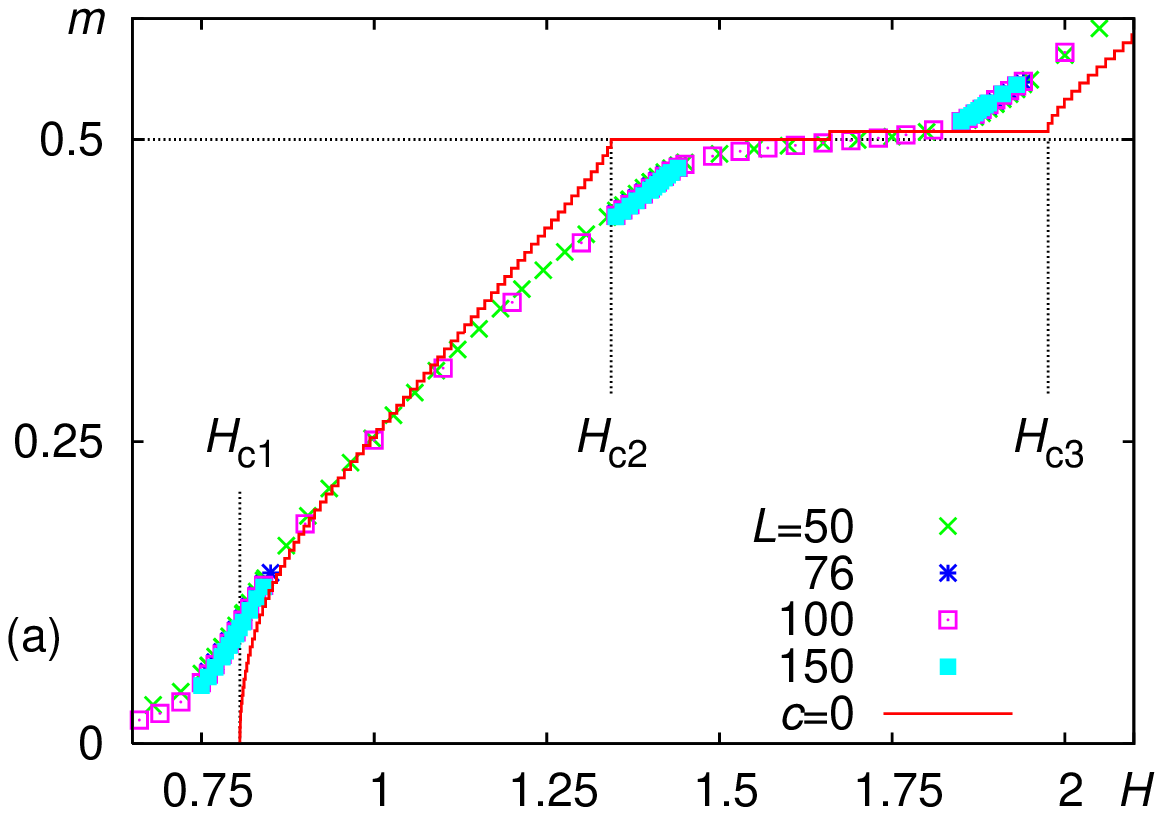}
\includegraphics[width=0.95\columnwidth]{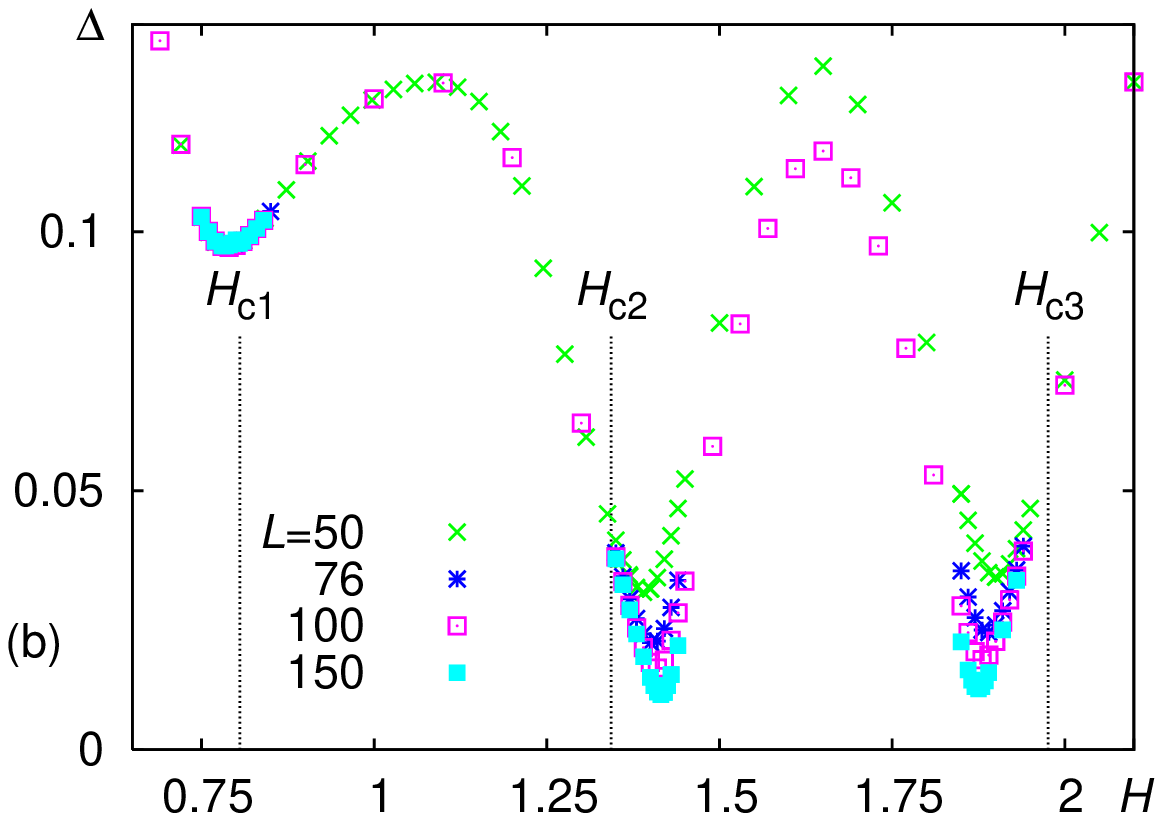}
\caption{(Color online)
Finite-size effects on magnetization (a) and energy gap (b)
in the presence of a small anisotropy $c = 0.03$.  The solid line
shows $m(H)$ for $c=0$ and $L=150$.  The dotted lines mark the
critical fields $H_{c1}$, $H_{c2}$, and $H_{c3}$.  $L$ is the length
of the ladders.}
\label{fig-anisotropy}
\end{figure}

Our numerical work is focused on the influence of the staggered field
$h_{0\pi}$.  (In what follows we drop the wavevector index.)  We choose
its amplitude to be proportional to the magnitude of the external
field, as happens with the Dzyaloshinskii-Moriya
interaction:\cite{Xia88,Oshikawa97b} $h = cH$.  The anisotropy coefficient
$c$ is varied between 0 and 0.1.  Fig.~\ref{fig-anisotropy} shows the
magnetization $m(H)$ and energy gap $\Delta(H)$ for $c = 0.03$ for
several lengths of the ladder $L$.

The introduction of the transverse field destroys the magnetization
plateaux: the spin projection $S^z$ is no longer conserved.  However,
the fate of the quantum critical points is different for the
transitions out of the integer ($H_{c1}$) and fractional ($H_{c2}$ and
$H_{c3}$) magnetization plateaux.  A complete lack of finite-size
effects near $H_{c1}$ is a good indication that the magnon
condensation point $H_{c1}$ has become a crossover.  In contrast, near
the points of spinon condensation, $H_{c2}$ and $H_{c3}$, the energy gap
$\Delta$ is still sensitive to the system size, indicating that the
critical points survive the introduction of anisotropy.

\subsection{Anisotropy and magnon condensation}
\label{section-anisotropic-Hc1}

\begin{figure}
\includegraphics[width=0.95\columnwidth]{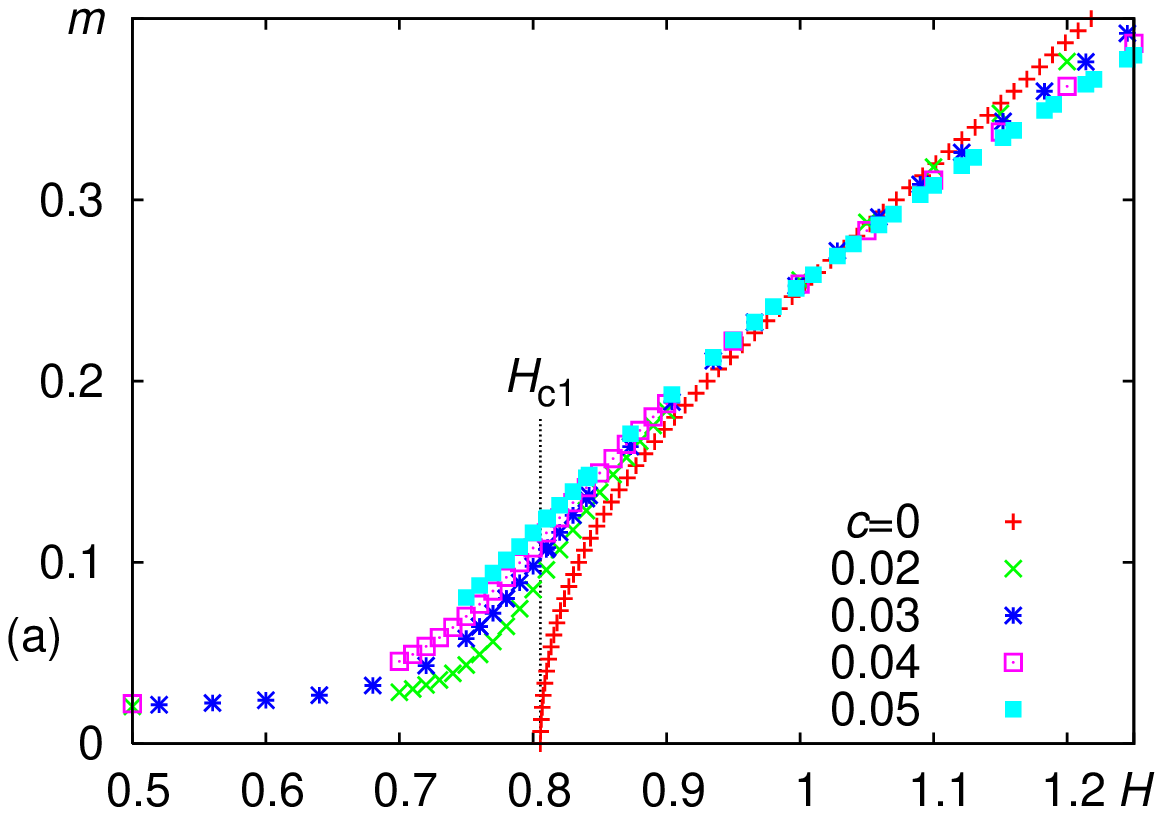}
\includegraphics[width=0.95\columnwidth]{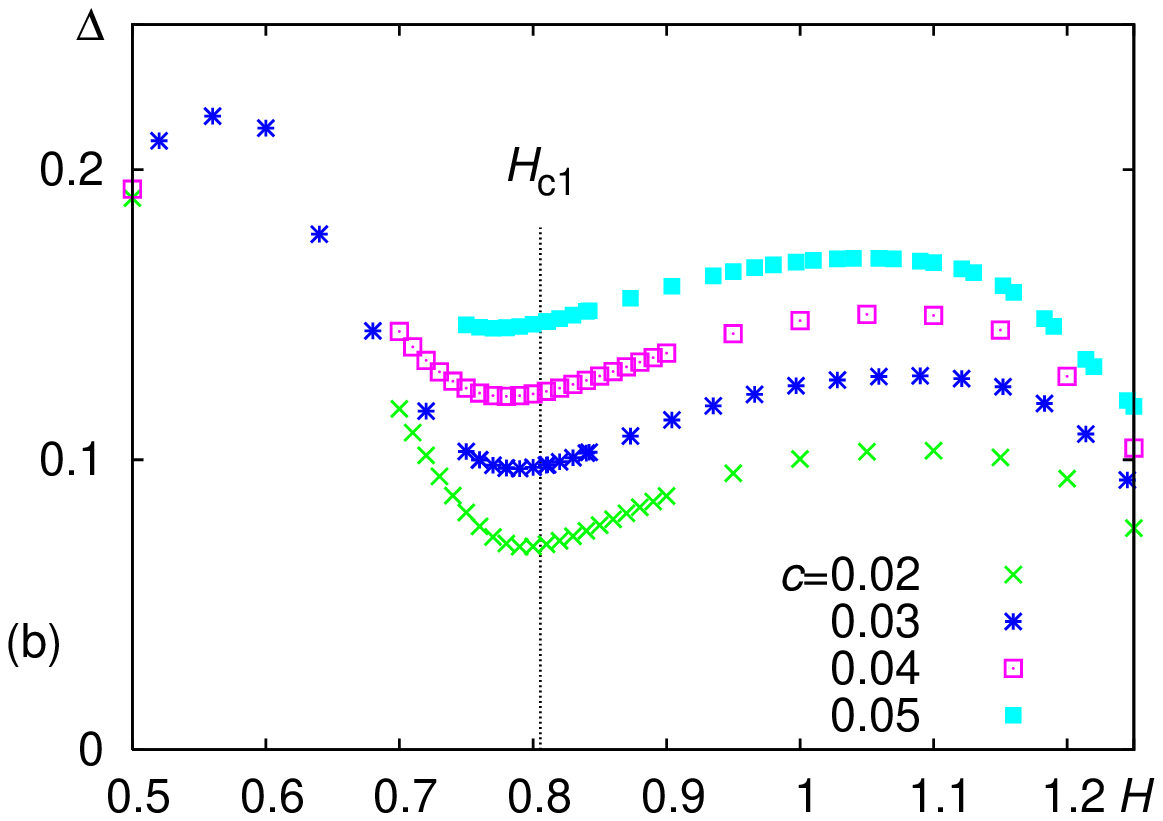}
\includegraphics[width=0.95\columnwidth]{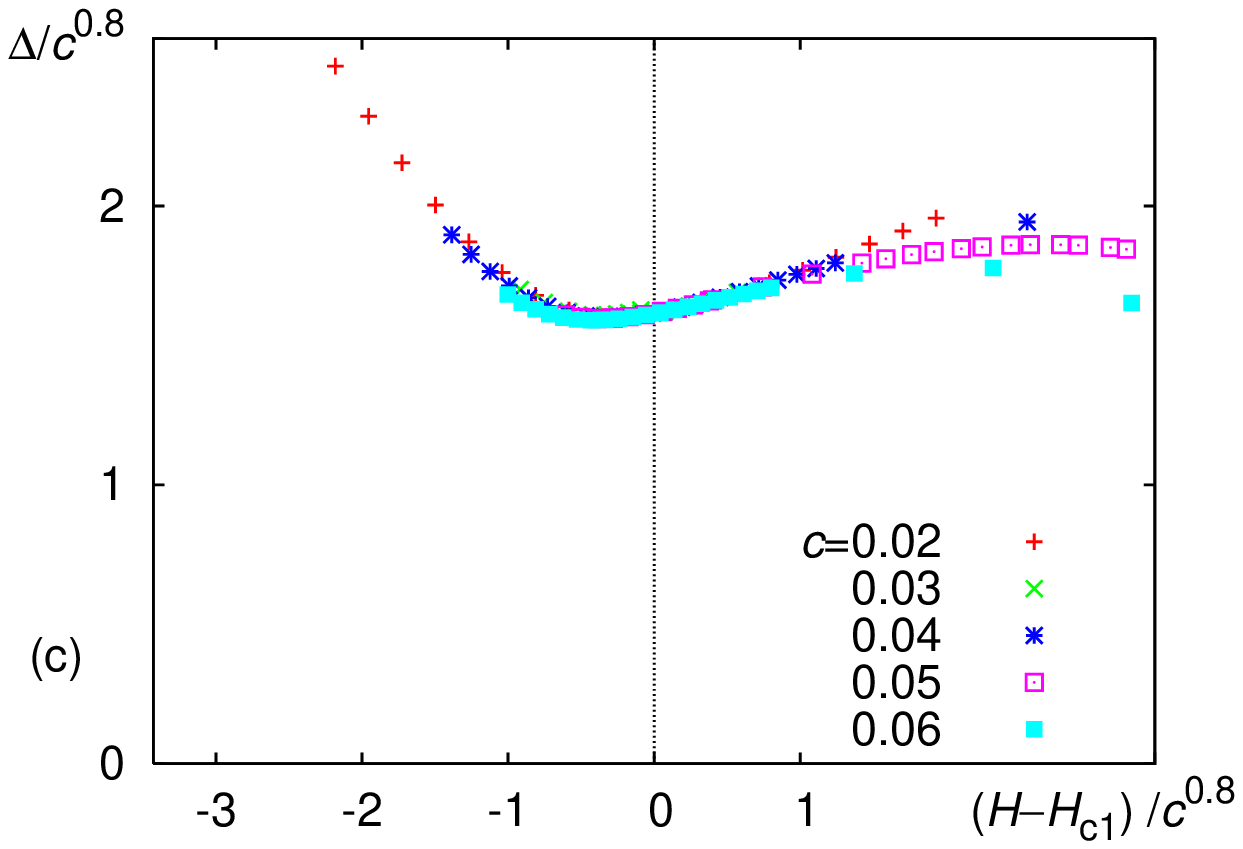}
\caption{(Color online)
Magnetization (a) and energy gap (b) near $H_{c1}$ for
several values of the anisotropy parameter $c$ (see text).  (c)
Scaling of the energy gap near $H_{c1}$.  The dotted line shows the
location of the critical field.}
\label{fig-Hc1}
\end{figure}

As noted above, the introduction of even a weak staggered field ($h =
0.03 H$) completely suppresses finite-size effects in ladders of
length $L = 50$ and more.  This is fully consistent with the scaling
theory of Bose condensation in the presence of a symmetry-breaking
transverse field.\cite{Fisher89,Fouet04} A finite transverse field $h
= cH$ generates a finite correlation length $\xi \propto \Delta^{-1}
\propto c^{-4/5}$.  Evidently,
for $c=0.03$ we have $\xi \lesssim 50$, so
that ladders with $L \geq 50$ are already in the thermodynamic limit.

The magnetization and energy gap for several values of anisotropy $c$
are shown in Figs.~\ref{fig-Hc1}(a) and (b).  The data are in
agreement with the scaling theory.\cite{Fouet04} As shown in
Fig.~\ref{fig-Hc1}(c), the energy gap obeys the scaling law
\begin{equation}
\Delta(H-H_{c1}, c) = c^{4/5} \Phi((H-H_{c1}) c^{-4/5}).
\end{equation}
Exactly at $H_{c1}$ the energy gap is proportional to $c^{4/5}$, while
the magnetization $m(H_{c1}) \propto c^{2/5}$.

\subsection{Anisotropy and spinon condensation}
\label{section-anisotropic-Hc2}

\begin{figure}
\begin{center}
\includegraphics[width=0.95\columnwidth]{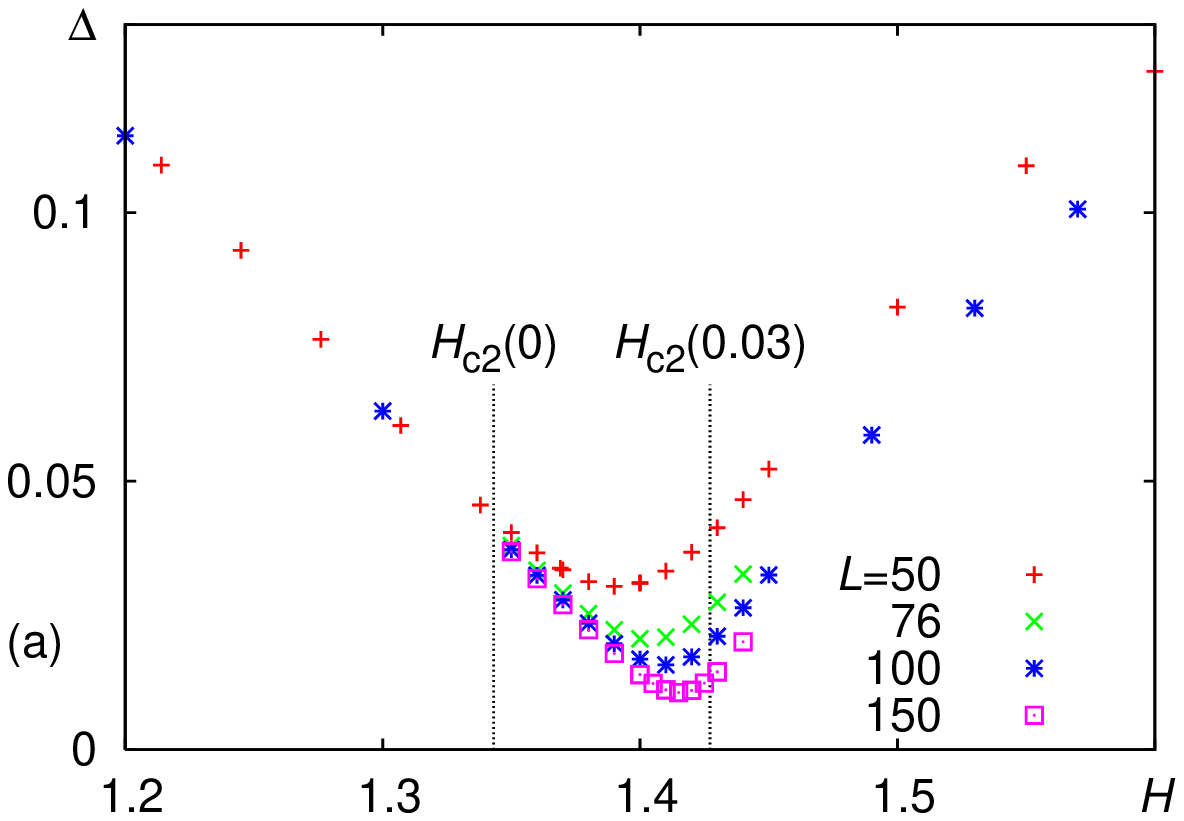}
\includegraphics[width=0.95\columnwidth]{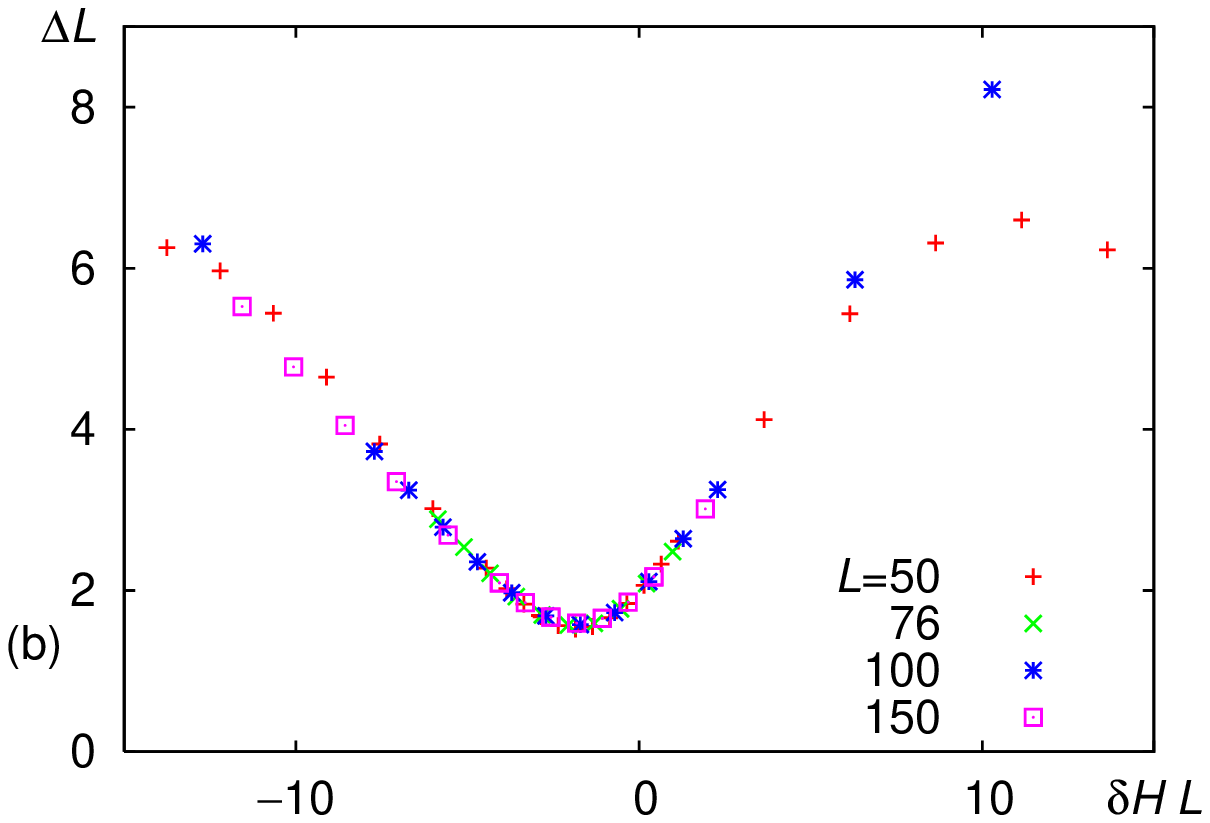}
\end{center}
\caption{(Color online)
Finite-size scaling of the energy gap near the critical field
$H_{c2}$ in the presence of finite anisotropy $c = 0.03$.  The dotted
line marks the location of $H_{c2}$ in the absence of anisotropy.}
\label{fig-Hc2-scaling}
\end{figure}

In contrast, for the same value of
anisotropy, strong finite-size effects are observed near the spinon
condensation points $H_{c2}$ and $H_{c3}$ (Fig.~\ref{fig-anisotropy}).
This indicates that the
quantum phase transitions survive the introduction of the staggered
transverse field $h_{0\pi}$.

The more robust nature of the spinon transitions can be traced to the
spontaneous breaking of a discrete lattice symmetry at the fractional
plateau: the two ground states (Fig.~\ref{fig-half}) break any
symmetry transformation  that exchanges even and odd lattice rungs.
The addition of the staggered field $h_{0\pi}$ keeps some of these
symmetries intact (e.g., the translation by one lattice spacing), so
that the Hamiltonian remains more symmetric than the ground states.
In other words, the $Z_2$ translational order remains in what used to
be the fractional plateau.  Restoration of the $Z_2$ symmetry at
$H_{c2}$ (and $H_{c3}$) requires a phase transition, whether the
transverse field $h_{0\pi}$ is present or not.

The difference can also be understood by looking at the effect of the
staggered field on the elementary excitations of the integer and
fractional plateaux.  By coupling to operators $D^+_n$ and $D^-_n$,
the staggered field adds or subtracts angular momentum 1.  As a
result, it creates or destroys single magnons (spin 1) at $H_{c1}$ and
$H_{c4}$ but pairs of spinons (spin 1/2) at $H_{c2}$ and $H_{c3}$.  A
plausible effective Hamiltonian for the spinons with low momenta near
the critical point in a weak transverse field would be
\begin{equation}
{\mathcal H}_\mathrm{xxz} = \sum_{p} [(p^2/2m - \mu)a^\dagger_p a_p
+ i v p (a^\dagger_p a^\dagger_{-p} - a_{-p} a_p)],
\label{eq-H-spinon-continuum}
\end{equation}
where $p$ is the spinon momentum, $\mu \propto H - H_{c2}$ is the
chemical potential, and $v \propto h_{0\pi}$ is a pairing field.
The energy gap behaves as follows:
\begin{equation}
\Delta = \left\{
\begin{array}{ll}
|\mu| & \mbox{ if } \mu < mv^2,\\
\sqrt{mv^2(2\mu+mv^2)} & \mbox{ if } \mu > mv^2.
\end{array}
\right.
\label{eq-gap-spinon-transverse}
\end{equation}

The spinon condensation in 1+1 dimensions is thus similar to the
commensurate--incommensurate transitions in two-dimensional
statistical mechanics.\cite{denNijs88} In the absence of the
transverse field $h_{0\pi}$, it belongs to the metal--insulator
(Pokrovsky--Talapov) universality class.  Switching on the field
converts the transition to the Ising universality class.

We have verified that, for a fixed anisotropy $c$, the finite-size
scaling of the energy gap is consistent with the Ising transition
in 1+1 dimensions:
\begin{equation}
\Delta(\delta H, L) = L^{-1} f(\delta H \, L),
\end{equation}
where $\delta H = H - H_{c2}(c)$.  Note that the critical field
depends on the anisotropy parameter $c$.  The scaling for $c=0.03$
is shown in Fig.~\ref{fig-Hc2-scaling}.

\begin{figure}
\begin{center}
\includegraphics[width=0.95\columnwidth]{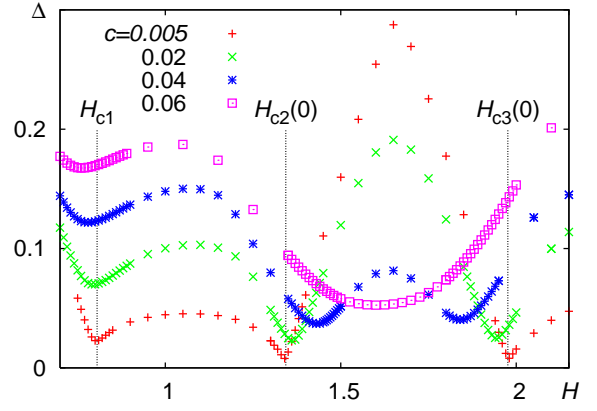}
\end{center}
\caption{(Color online)
Energy gap as a function of the field for several values of
anisotropy $c$ in a ladder of fixed length $L=50$.  The dotted lines
mark the critical fields $H_{c1}$, $H_{c2}$, and $H_{c3}$.}
\label{fig-gap-vs-H}
\end{figure}

As the anisotropy constant $c$ increases,
the critical fields $H_{c2}$ and $H_{c3}$ shift towards each other,
see Fig.~\ref{fig-gap-vs-H}.
The two fields merge and the ordered phase disappears at a
modest value of the anisotropy $c \approx 0.06$. Above this value of $c$,
the ground state is non-degenerate and does not break the translational
symmetry for any value of $H$.

Ideally, we would have liked to verify the universal scaling at the
quantum critical point $H=H_{c2}$, $c = 0$, $1/L = 0$, just like we
did at the magnon condensation point $H_{c1}$.  In this case, one
expects to observe a crossover from the Ising critical behavior to
that of the Pokrovsky--Talapov class.  As can be inferred from
Eq.~(\ref{eq-gap-spinon-transverse}), the energy gap is expected to
cross over from $|H-H_{c2}|$ on the gapped side to $|H-H_{c2}|^{1/2}$
in the formerly gapless region.  We have not been able to observe this
crossover.  This failure may be related to the narrowness of the
region where the spinons can be treated as noninteracting fermions
with a quadratic energy dispersion. (See
Sec.~\ref{section-isotropic-Hc2}.)  To observe the crossover, one
probably needs to work with a very small anisotropy $c$ (we went down
to $5 \times 10^{-4}$) and rather long ladders to avoid finite-size
effects.

\section{Discussion}
\label{sec-conclusion}

We have presented a model of a one-dimensional quantum antiferromagnet
in external magnetic field.  The system exhibits both integer and
fractional magnetization plateaux.  The quantum critical points at the
ends of an integer plateau are well-described by the Bose condensation
of magnons.  At low densities, the condensing magnons behave as
hard-core bosons or, alternatively, can be described as weakly interacting
fermions.  The introduction of a weak anisotropy fully breaking the
O(2) symmetry of the model replaces the quantum phase transition with
a crossover.  In contrast, a fractional magnetization plateau breaks a
$Z_2$ (Ising-like) symmetry of the lattice and ends in a condensation
of spinons -- domain walls in the $Z_2$ order parameter.  Magnetization
patterns at low spinon densities are explained by modeling the spinons
as free fermions.  The introduction of a weak anisotropy preserves the
$Z_2$ symmetry of the model.  As a result, the quantum phase
transition at the end of a fractional plateau survives.

This difference between integer and fractional plateaux is expected to
show up in the properties of quantum antiferromagnets in higher
dimensions as well. In that respect, SrCu$_2$(BO$_3$)$_2$, a physical
realization of the Shastry-Sutherland model, is a prominent candidate,
with plateaux at $m=1/8$, $m=1/4$ and $m=1/3$ that spontaneously
break the translational symmetry of the crystal.\cite{kageyama,kodama}
It is well established by now that there are significant
Dzyaloshinskii-Moriya interactions in that compound\cite{kodama2} and
that a gap persists in the region between the $m=0$ and $m=1/8$
plateaux, and closes (or has a deep minimum) before entering the
$m=1/8$ plateau. This behavior is reminiscent of the gap behavior we
found for the frustrated ladder.  Given the pecularities of the
triplet kinetic energy in the Shastry-Sutherland
model,\cite{miyahara,bendjama} the extension to that case of the ideas
developed in the present paper is far from trivial.

\section*{Acknowledgments}

We gratefully acknowledge discussions with F. D. M. Haldane, A.
L\"auchli, S. Miyahara, M. Oshikawa, O. Tretiakov, and M. Troyer.
This work was supported in part by the U.S. National Science
Foundation Grants No. DMR-0348679 and DMR-0412956, by the Swiss
National Fund, and by MaNEP.

\appendix

\section{The spin operators acting on a rung}
\label{ap-algebra}

There are four states on each rung, which decompose into a
singlet and a triplet under spin $\mathbf{S} = \mathbf{S}_1
+ \mathbf{S}_2$. We denote the singlet as $|s\rangle$, and the
$S_z=1,0,-1$ components of the triplet as $|+\rangle$, $|0\rangle$,
and $|-\rangle$ respectively. The
spin difference operator $\mathbf{D} = \mathbf{S}_1 - \mathbf{S}_2$
acts as follows:
\begin{eqnarray}
& D^z |s\rangle = |0\rangle,\qquad
D^z |0\rangle = & |s\rangle,\qquad
D^z |\pm\rangle = 0, \nonumber \\
& D^\pm|\mp\rangle = \pm \sqrt{2} |s\rangle, \qquad
& D^\pm |s\rangle = \pm \sqrt{2}|\pm\rangle,
\nonumber\\
&D^\pm |\pm \rangle = 0,& D^\pm |0\rangle = 0
\label{eq-D}
\end{eqnarray}
where $D^\pm = D^x \pm i D^y$.

\section{Spinon energy at $\mathcal{O}(\delta J^2)$}
\label{app-XXZ2}

To compute the second-order correction to the XXZ Hamiltonian, one
needs the action of ${\cal H}_2$ on states comprised of singlets and
$S_z=1$ triplets. It follows from Eq.\ (\ref{eq-D}) that ${\mathbf
D}_n\cdot {\mathbf D}_{n+1}$ acts non-trivially only on
nearest-neighbor pairs of singlets, giving for one pair
\begin{equation}\label{eq-pair}
{\cal H}_2 |ss\rangle = \frac{\delta J}{2} \left(|+-\rangle\, +\,
|-+\rangle\, -\, |00\rangle\right)
\end{equation}
One can now do second-order perturbation theory to compute the
correction to the XXZ Hamiltonian coming from annihilating and
creating pairs of triplets out of the adjacent singlet states. Since
these pairs are created between other magnons when the system is near
the $m=1/2$ plateau, the perturbation theory requires the diagonalization
of ${\cal H}_1$ on a length-4 chain of triplets. Calling the resultant
states $\psi_i$ and their ${\cal H}_1$ eigenvalues $\lambda_i$, one
finds that the shift in energy coming from having singlets on adjacent
sites is
\begin{equation}
E_{ss}^{(2)} =
\sum_i\frac{\abs{\bra{\psi_i}{\cal H}_2\ket{+ss+}}^2}{-2-\lambda_i}
= C(J) \, \delta J^2,
\end{equation}
where
\begin{equation}
C = \frac{1/20}{-2}
+ \frac{1/5}{-2+J}
+ \frac{(5-\sqrt{21})/20}{-2-\frac{-1+\sqrt{21}}{2}J}
+ \frac{(5+\sqrt{21})/20}{-2+\frac{1+\sqrt{21}}{2}J}.
\end{equation}
At $J=0.625$ we obtain $C=-2.026$.  The largest contribution comes
from the last term where the excited state lies rather close to the
spinon (an energy difference of 0.255).

\section{Magnon energy at $\mathcal{O}(\delta J^2)$}
\label{ap-mass}

In Sec.~\ref{section-isotropic-Hc1}, we have determined the inverse magnon
mass to leading order in $\delta J$.  Here we evaluate the
next-order correction in the limit of well-separated energy scales
(\ref{eq-scales}).  To this end, we consider the state of the ladder
with a single $S^z = +1$ triplet in the background of singlet states.
For $\delta J = 0$ any such state is an eigenstate of the Hamiltonian
$\mathcal{H}_0 + \mathcal{H}_1$ (\ref{eq-H}).  The first-order
correction in $\delta J$ induces the hopping of the magnon to the next
site with the amplitude $\delta J/2$, yielding an inverse mass
$1/m^* = |\delta J|$.

Second-order corrections to the magnon kinetic energy include hopping
through intermediate states with additional magnons.  An example of
such a process is shown in Fig.~\ref{fig-2nd}.  Such a process involves
the creation of a pair of magnons with total spin 0 next to the original
magnon and the subsequent destruction of a magnon pair by the perturbation
term $\mathcal{H}_2$.  Since both of these events have the amplitude
$\mathcal{O}(\delta J)$, we may treat the dynamics of the intermediate
3-magnon state using the zeroth-order Hamiltonian $\mathcal{H}_0 +
\mathcal{H}_1$.  At that level, the 3-magnon composite is immobile.
However, it has some internal dynamics: the spins of the individual
magnons $S^z_n$ may fluctuate.  We therefore first discuss the
internal dynamics of the composite.

A 3-magnon composite with the total spin $\Delta S^z = +1$ has 6
internal states:
\begin{equation}
\begin{array}{llllll}
|{-}{+}{+}\rangle,&
|{+}{-}{+}\rangle, &
|{+}{+}{-}\rangle, &
|{+}{0}{0}\rangle, &
|{0}{+}{0}\rangle, &
|{0}{0}{+}\rangle
\end{array}.
\end{equation}
In this basis, $\mathcal{H}_0 = 3-H$, while
\begin{equation}
\mathcal{H}_1 = J\left(
\begin{array}{cccccc}
0 & 0 & 0 & 0 & 0 & 1\\
0 &-2 & 0 & 1 & 0 & 1\\
0 & 0 & 0 & 1 & 0 & 0\\
0 & 1 & 1 & 0 & 1 & 0\\
0 & 0 & 0 & 1 & 0 & 1\\
1 & 1 & 0 & 0 & 1 & 0
\end{array}
\right).
\end{equation}
The lowest energy composite state is close to
that shown in Fig.~\ref{fig-2nd}(b) and has the energy $3 - H - 3J$.
The energy gap between the magnon and the 3-magnon composite is
$\Delta_1 = 2-3J$.  Our numerical studies were done on a system with
$J = 0.625$, which yields a gap $\Delta_1 = 0.125$, comparable to the
perturbation strength $\delta J = 0.15$.  It is therefore not
surprising that the leading-order result for the magnon mass was not
reliable.

To evaluate corrections at $\mathcal{O}(\delta J^2)$, we examine the
hybridization of the magnon $a_n$ with the symmetric 3-magnon
composites $c_{\gamma n}$ (where $\gamma = 1 \ldots 6$ is the internal
index):
\begin{eqnarray}
\mathcal{H}_\mathrm{magnon} &=&
\sum_{n} (\delta J/2)(a^\dagger_n a_{n-1} + \mathrm{H. c.})
\nonumber \\
&&+ \sum_{n} \sum_{\gamma = 1}^{6}
\Delta_\gamma c^\dagger_{\gamma n} c_{\gamma n}
\\
&&+ \sum_{n}
\sum_{\gamma = 1}^6
[c_{\gamma n}^\dagger(\lambda_\gamma a_{n+1} + \rho_\gamma a_{n-1})
+ \mathrm{H. c.}],
\nonumber
\end{eqnarray}
where $\lambda_{\gamma}$ and $\rho_{\gamma}$ are matrix elements
involved in creating a 3-magnon composite centered on the left or on
the right of the magnon:
\begin{equation}
\lambda_\gamma
= \langle s | c_{\gamma n} \, \mathcal{H}_2 \, a^\dagger_{n+1}| s \rangle,
\quad
\rho_\gamma
= \langle s | c_{\gamma n} \, \mathcal{H}_2 \, a^\dagger_{n-1}| s \rangle,
\end{equation}
where $|s\rangle$ is the all-singlet vacuum state.  Elimination of the
hybridization term to $\mathcal{O}(\delta J)$ via a unitary
transformation
\begin{equation}
c_{\gamma n} \, \mapsto \, c_{\gamma n}
- (\rho_\gamma a_{n-1} + \lambda_\gamma a_{n+1})/\Delta_\gamma
\end{equation}
generates an additional magnon hopping term at $\mathcal{O}(\delta J^2)$:
\begin{equation}
-\sum_{\gamma=1}^6 \frac{\rho_\gamma \lambda_\gamma}{\Delta_\gamma}
\sum_n a^\dagger_{n-1} a_{n+1} + \mathrm{H. c.}
\end{equation}
This yields the inverse magnon mass
\begin{eqnarray}
1/m^* &=& |\delta J| +
8\sum_{\gamma=1}^6 \frac{\rho_\gamma \lambda_\gamma}{\Delta_\gamma}
\\
&=& |\delta J|
+ 2\left(\frac{1}{6}-\frac{1}{2-J}+\frac{5}{3(2-3J)}\right) \delta J^2.
\nonumber
\end{eqnarray}
The calculation is simplified because the one-magnon state only
couples to three of the six composite states. For our choice of the
coupling constants, the second-order correction to the inverse mass
(0.575) is almost four times as large as the first-order term (0.15)
thanks to the small energy gap between one- and 3-magnon states.

The same unitary transformation allows us to determine the shift of
the magnon energy at this order:
\begin{equation}
\Delta \epsilon_{1,1}
= -\left(\frac{1}{6}+\frac{5}{3(2-3J)}\right)\delta J^2
+ E_\mathrm{Casimir}.
\label{eq-dE-magnon-2}
\end{equation}
The second term comes from the effect of the magnon on the vacuum.
The Hamiltonian term ${\cal H}_2$ creates virtual excitations in the
form of two triplets on adjacent rungs.  These virtual processes
shift the vacuum energy by
\begin{equation}
\Delta E_{0} = \sum_i
\frac{\abs{\bra{p_i}\mathcal{H}_2\ket{s}}^2}{-2+2J} =
\frac{-3L\delta J^2}{4(2-2J)} \; , \label{eq-vacuum-corr}
\end{equation}
where $\ket{p_i}$ is the intermediate state (\ref{eq-pair}) with two
triplets next to each other.  In the presence of a magnon the vacuum
fluctuations are suppressed in the immediate vicinity of the magnon.
The factor $L$ in Eq.~(\ref{eq-vacuum-corr}) is replaced with $L-4$,
increasing the energy of the 1-magnon state by the Casimir term in
Eq.~(\ref{eq-dE-magnon-2})
\begin{equation}
E_\mathrm{Casimir} = 3 \delta J^2/2(1-J).
\end{equation}

The second-order correction to the magnon condensation field is then
\begin{equation}
\Delta H_{c1} =
\Delta \epsilon_{1,1} = -9.5 \delta J^2 = -0.214.
\end{equation}
Again, because of the small energy gap to excited states, the second-
order correction to $H_{c1}$ exceeds the first-order one ($-0.15$) and
does not improve the agreement with the numerics.

\section{Envelope of staggered magnetization}
\label{ap-det}

In this Section we derive the expression for the envelope of the
staggered magnetization (\ref{eq-E-S}).  We work in the continuum
limit and treat spinons as noninteracting fermions in a box with
$0 < x < L$.  Since each spinon serves as a domain wall, the envelope
$E(x)$ is found by averaging the operator
\begin{equation}
\hat{E}(x) = \prod_{i=1}^r \mathrm{sgn}(x_i-x)
\end{equation}
over the (fully antisymmetric) wavefunction of $r$ spinons
\begin{equation}
\Psi(x_1 \ldots x_r) = \frac{1}{\sqrt{r!}}
\sum_{\{k\}} \varepsilon_{\{k\}}
\prod_{i=1}^r \psi_{k_i}(x_i),
\end{equation}
where $\varepsilon_{\{k\}} = \varepsilon_{k_1 \ldots k_r}$ is the
fully antisymmetric tensor with $\varepsilon_{12 \ldots r} = +1$.

The averaging yields
\begin{equation}
E(x) = \frac{1}{r!}\sum_{\{k\}} \sum_{\{l\}}
\varepsilon_{\{k\}} \varepsilon_{\{l\}}
\prod_{i=1}^r S_{k_i l_i}(x)
\end{equation}
with the matrix $S_{ij}(x)$ defined in Eq.~(\ref{eq-S}).  It can be
simplified by noting that $\varepsilon_{\{k\}} \varepsilon_{\{l\}} =
(-1)^P$, where $P$ is the permutation that maps $\{k\}$ into $\{l\}$.
Shifting from the sum over $\{k\}$ and $\{l\} = P(\{k\})$ to a sum
over $\{k\}$ and $P$ allows the sum over $\{k\}$ to be performed trivially
to obtain
\begin{equation}
E(x) = \sum_{P} (-1)^P \prod_{i=1}^r S_{i P_i}(x) = \det{S(x)}.
\end{equation}

\section{Reduction of staggered magnetization}

\label{app-reduction}

The domain wall magnetization envelope will be reduced by the presence
of quantum fluctuations. The leading order effect is the creation and
subsequent annihilation of a pair of domain walls due to the
interchange of a singlet and a triplet on neighboring sites.  The
domain wall Hamiltonian for the XXZ chain is\cite{Santos90}
\begin{eqnarray}
H &=& \sum_j \frac{J}{2} c_j^{\dagger} c_j
+\frac{\delta J}{2}(1-c_j^{\dagger} c_j)
\\
&&\times(c_{j+1}^{\dagger} c_{j-1}+c_{j+1}c_{j-1} + \mbox{ H.c.}),
\nonumber
\end{eqnarray}
where the $c$ operators obey fermionic commutation relations. We shall
ignore the quartic terms in this Hamiltonian in the approximation of
low spinon density.

Since these domain walls represent boundaries between the N\'eel states
with $S_n^z = (-1)^n/2$ and $(-1)^{n+1}/2$, the expectation value of
the spin on rung n in a state $\ket{\psi}$ is
\begin{equation}
\langle S_n^z \rangle = \frac{(-1)^n}{2}\bra{\psi}  E_n\ket{\psi}
\end{equation}
in addition to the spin carried by the spinons themselves. Here $
E_n = (-1)^{s(n)}$, where $s(n)$ is the number of spinons to the left of
rung $n$.

The spinon-number changing terms in the Hamiltonian can be eliminated
to lowest order in $\epsilon=\delta J/J$ by the unitary transformation
\begin{equation}
c_{n} \mapsto e^{-B}c_{n}e^{B}, \quad
B = \frac{\epsilon}{2} \sum_j c_{j+1}^{\dagger} c_{j-1}^{\dagger}
- \mbox{ H.c.}
\end{equation}

We can write the eigenstates of $H$ as
$\ket{\psi}=e^{B}\ket{\tilde{\psi}}$, where $\ket{\tilde{\psi}}$ is an
eigenstate of $e^{-B}He^{B}$ with definite
spinon number.  We now have that
\begin{widetext}
\begin{eqnarray}
S_n &=&\frac{(-1)^n}{2}\bra{\tilde{\psi}}e^{-B} E_n e^{B}\ket{\tilde{\psi}}
= \frac{(-1)^n}{2}
   \bra{\tilde{\psi}}
     E_n-[B,E_n]+\frac{1}{2}[B,[B,E_n]+\dots
   \ket{\tilde{\psi}}
\nonumber\\
&=&\frac{(-1)^n}{2} \bra{\tilde{\psi}}
E_n+\frac{\epsilon^2}{8}
 \left[
   \sum_j(c_{j+1}^{\dagger} c_{j-1}^{\dagger}+c_{j+1}c_{j-1}),
    \left[
      \sum_i(c_{i+1}^{\dagger} c_{i-1}^{\dagger}+c_{i+1}c_{i-1}),E_n
    \right]
 \right]
\ket{\tilde{\psi}}
\nonumber\\
&=& \frac{(-1)^n}{2}(1-\epsilon^2)
 \bra{\tilde{\psi}}E_n\ket{\tilde{\psi}}
+\mathcal{O}(1/L)
+\mathcal{O}(\epsilon^3).
\end{eqnarray}
\end{widetext}
Here we have used the relation
\begin{equation}
c_m E_n = \mathrm{sgn}(m-n) \,E_n c_m
\end{equation}
and its adjoint. Note that the first order term vanishes because the
operator $[B,E_n]$ includes a net change in the number of spinons. The
$O(1/L)$ piece arises from terms in the commutator proportional to the
spinon density operator.  The second order term is independent of the
number of spinons. Hence, the leading effect of the quantum
fluctuations is to reduce the value of the staggered magnetization by
a factor of $1-(\delta J/J)^2$.




\begin{thebibliography}{32}
\expandafter\ifx\csname natexlab\endcsname\relax\def\natexlab#1{#1}\fi
\expandafter\ifx\csname bibnamefont\endcsname\relax
  \def\bibnamefont#1{#1}\fi
\expandafter\ifx\csname bibfnamefont\endcsname\relax
  \def\bibfnamefont#1{#1}\fi
\expandafter\ifx\csname citenamefont\endcsname\relax
  \def\citenamefont#1{#1}\fi
\expandafter\ifx\csname url\endcsname\relax
  \def\url#1{\texttt{#1}}\fi
\expandafter\ifx\csname urlprefix\endcsname\relax\def\urlprefix{URL }\fi
\providecommand{\bibinfo}[2]{#2}
\providecommand{\eprint}[2][]{\url{#2}}

\bibitem[{\citenamefont{Sachdev}(2001)}]{Sachdev-book}
\bibinfo{author}{\bibfnamefont{S.}~\bibnamefont{Sachdev}},
  \emph{\bibinfo{title}{Quantum phase transitions}}
  (\bibinfo{publisher}{Cambridge University Press}, \bibinfo{year}{2001}).

\bibitem[{\citenamefont{Stewart}(1984)}]{Stewart84}
\bibinfo{author}{\bibfnamefont{G.~R.} \bibnamefont{Stewart}},
  \bibinfo{journal}{Rev. Mod. Phys.} \textbf{\bibinfo{volume}{56}},
  \bibinfo{pages}{755} (\bibinfo{year}{1984}).

\bibitem[{\citenamefont{Bitko et~al.}(1996)\citenamefont{Bitko, Rosenbaum, and
  Aeppli}}]{Bitko96}
\bibinfo{author}{\bibfnamefont{D.}~\bibnamefont{Bitko}},
  \bibinfo{author}{\bibfnamefont{T.~F.} \bibnamefont{Rosenbaum}},
  \bibnamefont{and} \bibinfo{author}{\bibfnamefont{G.}~\bibnamefont{Aeppli}},
  \bibinfo{journal}{Phys. Rev. Lett.} \textbf{\bibinfo{volume}{77}},
  \bibinfo{pages}{940} (\bibinfo{year}{1996}).

\bibitem[{\citenamefont{Greiner et~al.}(2002)\citenamefont{Greiner, Mandel,
  Esslinger, H\"{a}nsch, and Bloch}}]{Greiner02}
\bibinfo{author}{\bibfnamefont{M.}~\bibnamefont{Greiner}},
  \bibinfo{author}{\bibfnamefont{O.}~\bibnamefont{Mandel}},
  \bibinfo{author}{\bibfnamefont{T.}~\bibnamefont{Esslinger}},
  \bibinfo{author}{\bibfnamefont{T.~W.} \bibnamefont{H\"{a}nsch}},
  \bibnamefont{and} \bibinfo{author}{\bibfnamefont{I.}~\bibnamefont{Bloch}},
  \bibinfo{journal}{Nature} \textbf{\bibinfo{volume}{415}}, \bibinfo{pages}{39}
  (\bibinfo{year}{2002}).

\bibitem[{\citenamefont{Affleck}(1990)}]{Affleck90}
\bibinfo{author}{\bibfnamefont{I.}~\bibnamefont{Affleck}},
  \bibinfo{journal}{Phys. Rev. B}
  \textbf{\bibinfo{volume}{41}},
  \bibinfo{pages}{6697} (\bibinfo{year}{1990});
  \textbf{\bibinfo{volume}{43}},
  \bibinfo{pages}{3215} (\bibinfo{year}{1991}).

\bibitem[{\citenamefont{Giamarchi and Tsvelik}(1999)}]{giamarchi}
\bibinfo{author}{\bibfnamefont{T.}~\bibnamefont{Giamarchi}} \bibnamefont{and}
  \bibinfo{author}{\bibfnamefont{A.~M.} \bibnamefont{Tsvelik}},
  \bibinfo{journal}{Phys. Rev. B} \textbf{\bibinfo{volume}{59}},
  \bibinfo{pages}{11398} (\bibinfo{year}{1999}).

\bibitem[{\citenamefont{Fisher et~al.}(1989)\citenamefont{Fisher, Weichman,
  Grinstein, and Fisher}}]{Fisher89}
\bibinfo{author}{\bibfnamefont{M.~P.~A.} \bibnamefont{Fisher}},
  \bibinfo{author}{\bibfnamefont{P.~B.} \bibnamefont{Weichman}},
  \bibinfo{author}{\bibfnamefont{G.}~\bibnamefont{Grinstein}},
  \bibnamefont{and} \bibinfo{author}{\bibfnamefont{D.~S.}
  \bibnamefont{Fisher}}, \bibinfo{journal}{Phys. Rev. B}
  \textbf{\bibinfo{volume}{40}}, \bibinfo{pages}{546} (\bibinfo{year}{1989}).

\bibitem[{\citenamefont{Oshikawa et~al.}(1997)\citenamefont{Oshikawa, Yamanaka,
  and Affleck}}]{Oshikawa97a}
\bibinfo{author}{\bibfnamefont{M.}~\bibnamefont{Oshikawa}},
  \bibinfo{author}{\bibfnamefont{M.}~\bibnamefont{Yamanaka}}, \bibnamefont{and}
  \bibinfo{author}{\bibfnamefont{I.}~\bibnamefont{Affleck}},
  \bibinfo{journal}{Phys. Rev. Lett.} \textbf{\bibinfo{volume}{78}},
  \bibinfo{pages}{1984} (\bibinfo{year}{1997}).

\bibitem[{\citenamefont{Fendley et~al.}(2004)\citenamefont{Fendley, Sengupta,
  and Sachdev}}]{Fendley04}
\bibinfo{author}{\bibfnamefont{P.}~\bibnamefont{Fendley}},
  \bibinfo{author}{\bibfnamefont{K.}~\bibnamefont{Sengupta}}, \bibnamefont{and}
  \bibinfo{author}{\bibfnamefont{S.}~\bibnamefont{Sachdev}},
  \bibinfo{journal}{Physical Review B} \textbf{\bibinfo{volume}{69}},
  \bibinfo{pages}{075106} (\bibinfo{year}{2004}).

\bibitem[{\citenamefont{Lecheminant and Orignac}(2004)}]{Lecheminant04}
\bibinfo{author}{\bibfnamefont{P.}~\bibnamefont{Lecheminant}} \bibnamefont{and}
  \bibinfo{author}{\bibfnamefont{E.}~\bibnamefont{Orignac}},
  \bibinfo{journal}{Physical Review B} \textbf{\bibinfo{volume}{69}},
  \bibinfo{pages}{174409} (\bibinfo{year}{2004}).

\bibitem[{\citenamefont{Essler and Affleck}(2004)}]{Essler04}
\bibinfo{author}{\bibfnamefont{F.~H.~L.} \bibnamefont{Essler}}
  \bibnamefont{and} \bibinfo{author}{\bibfnamefont{I.}~\bibnamefont{Affleck}},
  \bibinfo{journal}{J. Stat. Mech.} \textbf{\bibinfo{volume}{P12006}}
  (\bibinfo{year}{2004}).

\bibitem[{\citenamefont{Xia and Riseborough}(1988)}]{Xia88}
\bibinfo{author}{\bibfnamefont{Q.}~\bibnamefont{Xia}} \bibnamefont{and}
  \bibinfo{author}{\bibfnamefont{P.~S.} \bibnamefont{Riseborough}},
  \bibinfo{journal}{J. Appl. Phys.} \textbf{\bibinfo{volume}{63}},
  \bibinfo{pages}{4141} (\bibinfo{year}{1988}).

\bibitem[{\citenamefont{Oshikawa and Affleck}(1997)}]{Oshikawa97b}
\bibinfo{author}{\bibfnamefont{M.}~\bibnamefont{Oshikawa}} \bibnamefont{and}
  \bibinfo{author}{\bibfnamefont{I.}~\bibnamefont{Affleck}},
  \bibinfo{journal}{Phys. Rev. Lett.} \textbf{\bibinfo{volume}{79}},
  \bibinfo{pages}{2883} (\bibinfo{year}{1997}).

\bibitem[{\citenamefont{Mila}(1998)}]{Mila98}
\bibinfo{author}{\bibfnamefont{F.}~\bibnamefont{Mila}}, \bibinfo{journal}{Eur.
  Phys. J. B} \textbf{\bibinfo{volume}{6}}, \bibinfo{pages}{201}
  (\bibinfo{year}{1998}).

\bibitem[{\citenamefont{Cabra et~al.}(1997)\citenamefont{Cabra, Honecker, and
  Pujol}}]{Cabra97}
\bibinfo{author}{\bibfnamefont{D.~C.} \bibnamefont{Cabra}},
  \bibinfo{author}{\bibfnamefont{A.}~\bibnamefont{Honecker}}, \bibnamefont{and}
  \bibinfo{author}{\bibfnamefont{P.}~\bibnamefont{Pujol}},
  \bibinfo{journal}{Phys. Rev. Lett.} \textbf{\bibinfo{volume}{79}},
  \bibinfo{pages}{5126} (\bibinfo{year}{1997}).

\bibitem[{\citenamefont{Honecker et~al.}(2000)\citenamefont{Honecker, Mila, and
  Troyer}}]{Honecker00}
\bibinfo{author}{\bibfnamefont{A.}~\bibnamefont{Honecker}},
  \bibinfo{author}{\bibfnamefont{F.}~\bibnamefont{Mila}}, \bibnamefont{and}
  \bibinfo{author}{\bibfnamefont{M.}~\bibnamefont{Troyer}},
  \bibinfo{journal}{Eur. Phys. J. B} \textbf{\bibinfo{volume}{15}},
  \bibinfo{pages}{227} (\bibinfo{year}{2000}).

\bibitem[{\citenamefont{Yang and Yang}(1967)}]{YangIII}
\bibinfo{author}{\bibfnamefont{C.~N.} \bibnamefont{Yang}} \bibnamefont{and}
  \bibinfo{author}{\bibfnamefont{C.~P.} \bibnamefont{Yang}},
  \bibinfo{journal}{Phys. Rev.} \textbf{\bibinfo{volume}{151}},
  \bibinfo{pages}{258} (\bibinfo{year}{1967}).

\bibitem[{\citenamefont{Haldane}(1981)}]{Haldane81}
\bibinfo{author}{\bibfnamefont{F.~D.~M.} \bibnamefont{Haldane}},
  \bibinfo{journal}{J. Phys. C} \textbf{\bibinfo{volume}{14}},
  \bibinfo{pages}{2585} (\bibinfo{year}{1981}).

\bibitem[{\citenamefont{Fowler and Puga}(1978)}]{Fowler78}
\bibinfo{author}{\bibfnamefont{M.}~\bibnamefont{Fowler}} \bibnamefont{and}
  \bibinfo{author}{\bibfnamefont{M.~W.} \bibnamefont{Puga}},
  \bibinfo{journal}{Phys. Rev. B} \textbf{\bibinfo{volume}{18}},
  \bibinfo{pages}{421} (\bibinfo{year}{1978}).

\bibitem[{\citenamefont{White}(1992)}]{White92}
\bibinfo{author}{\bibfnamefont{S.~R.} \bibnamefont{White}},
  \bibinfo{journal}{Phys. Rev. Lett.} \textbf{\bibinfo{volume}{69}},
  \bibinfo{pages}{2863} (\bibinfo{year}{1992}).

\bibitem[{\citenamefont{Noack and Manmana}(2005)}]{Noack05}
\bibinfo{author}{\bibfnamefont{R.~M.} \bibnamefont{Noack}} \bibnamefont{and}
  \bibinfo{author}{\bibfnamefont{S.~R.} \bibnamefont{Manmana}},
  \bibinfo{journal}{AIP Conf. Proc.} \textbf{\bibinfo{volume}{789}},
  \bibinfo{pages}{93} (\bibinfo{year}{2005}).

\bibitem[{\citenamefont{Haldane}(1980)}]{Haldane80}
\bibinfo{author}{\bibfnamefont{F.~D.~M.} \bibnamefont{Haldane}},
  \bibinfo{journal}{Phys. Rev. Lett.} \textbf{\bibinfo{volume}{45}},
  \bibinfo{pages}{1358} (\bibinfo{year}{1980}).

\bibitem[{\citenamefont{White et~al.}(2002)\citenamefont{White, Affleck, and
  Scalapino}}]{White02}
\bibinfo{author}{\bibfnamefont{S.~R.} \bibnamefont{White}},
  \bibinfo{author}{\bibfnamefont{I.}~\bibnamefont{Affleck}}, \bibnamefont{and}
  \bibinfo{author}{\bibfnamefont{D.~J.} \bibnamefont{Scalapino}},
  \bibinfo{journal}{Phys. Rev. B} \textbf{\bibinfo{volume}{65}},
  \bibinfo{pages}{165122} (\bibinfo{year}{2002}).

\bibitem[{\citenamefont{Sachdev}(1992)}]{Sachdev92}
\bibinfo{author}{\bibfnamefont{S.}~\bibnamefont{Sachdev}},
  \bibinfo{journal}{Phys. Rev. B} \textbf{\bibinfo{volume}{45}},
  \bibinfo{pages}{12377} (\bibinfo{year}{1992}).

\bibitem[{\citenamefont{Fouet et~al.}(2004)\citenamefont{Fouet, Tchernyshyov,
  and Mila}}]{Fouet04}
\bibinfo{author}{\bibfnamefont{J.-B.} \bibnamefont{Fouet}},
  \bibinfo{author}{\bibfnamefont{O.}~\bibnamefont{Tchernyshyov}},
  \bibnamefont{and} \bibinfo{author}{\bibfnamefont{F.}~\bibnamefont{Mila}},
  \bibinfo{journal}{Phys. Rev. B} \textbf{\bibinfo{volume}{70}},
  \bibinfo{pages}{174427} (\bibinfo{year}{2004}).

\bibitem[{\citenamefont{den Nijs}(1988)}]{denNijs88}
\bibinfo{author}{\bibfnamefont{M.}~\bibnamefont{den Nijs}}, in
  \emph{\bibinfo{booktitle}{Phase Transitions and Critical Phenomena}}, edited
  by \bibinfo{editor}{\bibfnamefont{C.}~\bibnamefont{Domb}} \bibnamefont{and}
  \bibinfo{editor}{\bibfnamefont{J.~L.} \bibnamefont{Lebowitz}}
  (\bibinfo{publisher}{Academic Press}, \bibinfo{year}{1988}),
  vol.~\bibinfo{volume}{12}.

\bibitem[{\citenamefont{Kageyama et~al.}(1999)\citenamefont{Kageyama,
  Yoshimura, Stern, Mushnikov, Onizuka, Kato, Kosuge, Slichter, Goto, and
  Ueda}}]{kageyama}
\bibinfo{author}{\bibfnamefont{H.}~\bibnamefont{Kageyama}},
  \bibinfo{author}{\bibfnamefont{K.}~\bibnamefont{Yoshimura}},
  \bibinfo{author}{\bibfnamefont{R.}~\bibnamefont{Stern}},
  \bibinfo{author}{\bibfnamefont{N.~V.} \bibnamefont{Mushnikov}},
  \bibinfo{author}{\bibfnamefont{K.}~\bibnamefont{Onizuka}},
  \bibinfo{author}{\bibfnamefont{M.}~\bibnamefont{Kato}},
  \bibinfo{author}{\bibfnamefont{K.}~\bibnamefont{Kosuge}},
  \bibinfo{author}{\bibfnamefont{C.~P.} \bibnamefont{Slichter}},
  \bibinfo{author}{\bibfnamefont{T.}~\bibnamefont{Goto}}, \bibnamefont{and}
  \bibinfo{author}{\bibfnamefont{Y.}~\bibnamefont{Ueda}},
  \bibinfo{journal}{Phys. Rev. Lett.} \textbf{\bibinfo{volume}{82}},
  \bibinfo{pages}{3168} (\bibinfo{year}{1999}).

\bibitem[{\citenamefont{Kodama et~al.}(2002)\citenamefont{Kodama, Takigawa,
  Horvatic, Berthier, Kageyama, Ueda, Miyahara, Becca, and Mila}}]{kodama}
\bibinfo{author}{\bibfnamefont{K.}~\bibnamefont{Kodama}},
  \bibinfo{author}{\bibfnamefont{M.}~\bibnamefont{Takigawa}},
  \bibinfo{author}{\bibfnamefont{M.}~\bibnamefont{Horvatic}},
  \bibinfo{author}{\bibfnamefont{C.}~\bibnamefont{Berthier}},
  \bibinfo{author}{\bibfnamefont{H.}~\bibnamefont{Kageyama}},
  \bibinfo{author}{\bibfnamefont{Y.}~\bibnamefont{Ueda}},
  \bibinfo{author}{\bibfnamefont{S.}~\bibnamefont{Miyahara}},
  \bibinfo{author}{\bibfnamefont{F.}~\bibnamefont{Becca}}, \bibnamefont{and}
  \bibinfo{author}{\bibfnamefont{F.}~\bibnamefont{Mila}},
  \bibinfo{journal}{Science} \textbf{\bibinfo{volume}{298}},
  \bibinfo{pages}{395} (\bibinfo{year}{2002}).

\bibitem[{\citenamefont{Kodama et~al.}(2005)\citenamefont{Kodama, Miyahara,
  Takigawa, Horvatic, Berthier, Mila, Kageyama, and Ueda}}]{kodama2}
\bibinfo{author}{\bibfnamefont{K.}~\bibnamefont{Kodama}},
  \bibinfo{author}{\bibfnamefont{S.}~\bibnamefont{Miyahara}},
  \bibinfo{author}{\bibfnamefont{M.}~\bibnamefont{Takigawa}},
  \bibinfo{author}{\bibfnamefont{M.}~\bibnamefont{Horvatic}},
  \bibinfo{author}{\bibfnamefont{C.}~\bibnamefont{Berthier}},
  \bibinfo{author}{\bibfnamefont{F.}~\bibnamefont{Mila}},
  \bibinfo{author}{\bibfnamefont{K.}~\bibnamefont{Kageyama}}, \bibnamefont{and}
  \bibinfo{author}{\bibfnamefont{Y.}~\bibnamefont{Ueda}}, \bibinfo{journal}{J.
  Phys.: Condens. Mat} \textbf{\bibinfo{volume}{17}}, \bibinfo{pages}{L61}
  (\bibinfo{year}{2005}).

\bibitem[{\citenamefont{Miyahara and Ueda}(2003)}]{miyahara}
\bibinfo{author}{\bibfnamefont{S.}~\bibnamefont{Miyahara}} \bibnamefont{and}
  \bibinfo{author}{\bibfnamefont{K.}~\bibnamefont{Ueda}}, \bibinfo{journal}{J.
  Phys.: Cond. Matt.} \textbf{\bibinfo{volume}{15}}, \bibinfo{pages}{R327}
  (\bibinfo{year}{2003}).

\bibitem[{\citenamefont{Bendjama et~al.}(2005)\citenamefont{Bendjama, Kumar,
  and Mila}}]{bendjama}
\bibinfo{author}{\bibfnamefont{R.}~\bibnamefont{Bendjama}},
  \bibinfo{author}{\bibfnamefont{B.}~\bibnamefont{Kumar}}, \bibnamefont{and}
  \bibinfo{author}{\bibfnamefont{F.}~\bibnamefont{Mila}},
  \bibinfo{journal}{Phys. Rev. Lett.} \textbf{\bibinfo{volume}{95}},
  \bibinfo{pages}{110406} (\bibinfo{year}{2005}).

\bibitem[{\citenamefont{G\'{o}mez-Santos}(1990)}]{Santos90}
\bibinfo{author}{\bibfnamefont{G.}~\bibnamefont{G\'{o}mez-Santos}},
  \bibinfo{journal}{Phys. Rev. B} \textbf{\bibinfo{volume}{41}},
  \bibinfo{pages}{6788} (\bibinfo{year}{1990}).

\end{thebibliography}
\end{document}